\documentclass[useAMS,usenatbib]{mn2e}
\usepackage{graphicx}

\def\la{\raise.5ex\hbox{$<$}\kern-.8em\lower 1mm\hbox{$\sim$}}
\def\ma{\raise.5ex\hbox{$>$}\kern-.8em\lower 1mm\hbox{$\sim$}}

\def\msol{M$_{\odot}$ }

\def\kms{$\rm km\, s^{-1}$}
\def\cm3{$\rm cm^{-3}$}
\def\Ts{$\rm T_{*}$~}
\def\Vs{$\rm V_{s}$~}
\def\n0{$\rm n_{0}$}
\def\B0{$\rm B_{0}$}

\def\Ne{$\rm N_{e}$}

\def\erg{$\rm erg\, cm^{-2}\, s^{-1}$}

\def\L12{L$_{12\mu m}$~}
\def\F12{F$_{12\mu m}$~}

\def\Hb{H${\beta}$~}
\def\Ha{H${\alpha}$}
\def\Hg{H${\gamma}$}

\def\RO3{R$_{[OIII]}$}

\title[N/O in the host galaxies of SN, GRB, AGN, SB and HII regions]{N/O abundance ratios 
in   gamma-ray burst and  supernova host galaxies at z$<$4. Comparison with AGN, starburst and HII regions
}

\author[M. Contini]{M. Contini 
\\
School of Physics and Astronomy, Tel Aviv University, Tel Aviv
69978, Israel \\
}

\begin{document}

%\date{Accepted: Received ; in original form 2010 month day}

\pagerange{\pageref{firstpage}--\pageref{lastpage}} \pubyear{2009}

\maketitle

\label{firstpage}

\begin{abstract}
The distribution of the N/O element abundance ratios calculated by the detailed modelling of different 
galaxy  spectra at z$<$4 is investigated. Supernova (SN) and long gamma-ray-burst (LGRB) host galaxies
cover  different redshift domains.  N/O ratios  in SN hosts increase due to
secondary N production towards low z (0.01)    accompanying the  growing  trend of active
 galaxies (AGN, LINER).   N/O ratios in LGRB hosts decrease rapidly between  z$>$1 and z$\sim$0.1 
following the N/H  trend and reach   the 
characteristic N/O ratios   calculated  for the  HII regions  in local and  nearby galaxies.
The few short period GRB (SGRB)  hosts included in the galaxy sample show N/H $\leq$ 0.04 solar  and O/H solar.
They seem to   continue the   low bound N/H trend of SN hosts at z$<$0.3.
The distribution of N/O  as function of metallicity for SN and LGRB hosts
is  compared with   star chemical evolution models.
The results show that  several  LGRB hosts can be explained by star multi-bursting  models  when 
12+log(O/H) $<$8.5,  while  some objects follow the trend of  continuous star formation models.
N/O in SN hosts at log(O/H)+12 $<$8.5 are   not well explained by stellar chemical evolution models  calculated for starburst 
galaxies.
At 12+log(O/H) $>$ 8.5 many different objects are nested close to O/H solar  with  N/O
ranging between the maximum  corresponding to starburst galaxies  and AGN and  the minimum  corresponding 
to HII regions and SGRB.
 
\end{abstract}


\section{Introduction}

Metallicity is one of the main  factors which affects the evolution of massive stars as well 
as their explosive death (Vergani et al 2011, Piranomonte et al 2015, Sollerman et al 2005, Woosley 1993, etc).
In previous papers (e.g. Contini 2017, 2016a and references therein) we have studied the characterisctic 
physical parameters and the element abundances
 of SN and GRB host galaxies  at  intermediate-high  redshifts, z$<$4. 
The sample of galaxies that was  assembled 
in order to study the role of the main physical and chemical parameters
 dominating in each object  (see e.g. Contini 2014 and references therein), 
accounts for many  types,  AGN, LINERs, in particular mergers, starbursts (SB), SN,  GRB hosts, etc.
 Lyman $\alpha$ and UV line emitting galaxies (Contini 2015) were included to expand to the UV domain the  line wavelengths  
from different elements  and ionization levels  observable  throughout the redshift ranges. 
We have  distinguished the host galaxies of SN (Contini 2016a)
in  different types (Ic, Ib, Ibc, IcBL, IIb, etc., Sanders et al. 2012; SLSNR, SLSNI, SLSN, Leloudas et al, 2015) 
and  GRB hosts (Contini 2016a, 2017) in long and short period on the basis of data 
by  de Ugarte Postigo et al (2014),
Kr\"{u}hler et al. (2015), Savaglio et al  (2009), Han et al (2010), etc.
In objects such as  AGN  and  SB,  the  characteristics of the radiation flux 
photoionizing and heating the gas are of primary importance to understand the distribution of the
gas physical conditions throughout the host galaxy. 
However, the element abundances are not less significant, because
metallicities   are related to the  star formation history.
SN and  GRB host galaxies  are also contaminated by  the explosion products.
The mutual  influence of SN,  GRB and their hosts have been discussed by Contini (2016a)
who suggested that  GRB hosts are less  affected  than  the SN ones.  

Metallicities in galaxies are revealed by the spectra.
In previous works we compared  N/H and O/H  calculated for SN and GRB
with those of SB,  AGN and LINER  galaxies as function of the redshift.
We performed our investigations  interpreting  line and continuum spectra which contained enough data
to constrain  the models.
 There are a few observed spectra at z$>$4 that have been modelled in detail.
Therefore,
 in this paper we   investigate the  evolution of  N/O  abundance ratios  with z in different 
 SN and GRB host galaxies focusing on  relatively  low redshift domain.
N/O ratios are directly linked to metallicity variations and to the formation of stars.
The mass of stars in a galaxy is related to its metallicity evolution (Tremonti et al. 2004).
 We  try to  explain 
the behaviour of N/O as function of z   by the theories of P\'{e}rez-Montero et al (20013), 
Henry et al. (2000), Mouhcine \& Contini (2002, hereafter MC02), etc.
Oxygen is produced only by massive stars. 
Analyzing  the behaviour of N/O  abundance ratios as a function of metallicity  in terms of O/H
 for Galactic and extragalactic H II region samples, Henry et al (2000)
found that massive stars (M $>$ 8 \msol) dominate the production of carbon, while intermediate-mass
stars between 4 and 8 \msol dominate nitrogen production.
They claim that N production is primary at low metallicities, but when 12 + log(O/H)  exceeds 8.3,
 secondary  N production occurs, increasing  at a faster rate than O, 
steeper than  expected for a secondary element. 
Moreover, they suggest that  the distribution of  N/O data is caused
by intermittent increases in nitrogen caused by local contamination by Wolf-Rayet stars or luminous
blue variables. In addition, the effect of inflow of gas into galactic systems on secondary production
of nitrogen from carbon may introduce some scatter into N/O ratios at high metallicities.
The evolution of  N/O  depends on internal star processes such as nucleosynthesis etc.
Massive stars ($\sim$ 8-20 \msol) are generally responsible  for SN explosions and GRB  in galaxies.
P\'{e}rez-Montero et al claim that N/O  is probably a chemical evolution indicator 
of single starbursts (Edmunds \& Pagel 1978, Pilyugin et al 2003). The relation of N (which is a secondary 
element) with O (which is a primary element) is relatively independent of the chemodynamical effects
such as outflows and inflows of gas from regions where star formation and element synthesis is occurring. 
They
can be only used to explain  structures of modified  metallicity 
throughout the galaxy  (Edmunds 1990). The chemical evolution of galaxies that undergo an episode of massive 
and rapid accretion
of metal poor gas (K\"{o}ppen \& Hensler 2005) provides a large scatter in the N/O versus O/H diagram.
P\'{e}rez-Montero \& Contini (2009) found an increase of N/O with stellar mass when most of the galaxies 
lie in the metallicity range for the production of the secondary N.
Element  abundances in  SN and GRB hosts as well as in the other galaxies are the product of  
intermingled physical  events such as SN feedback, gas inflow and outflow, 
mergers and interactions (P\'{e}rez-Montero et al 2013).
 Summarizing, nitrogen is produced as a primary element in massive stars (core collapse) supernovae,
then as a secondary element in WN stars and later in AGB stars.
Therefore it is  important to  obtain  the N/O results  from the detailed analysis of the line
spectra, in particular of the  N and O characteristic lines.  

The lines  are emitted from  gas in the  physical and chemical  conditions  
which are determined
throughout the galaxy  by  radiation sources and  dynamical effects. 
Notice that the emitting gas is not at a constant temperature and
density, but behind the ionization fronts  created by photoionization and
 downstream  of shock fronts when collisional ionization and heating dominate,
the gas recombines  following the cooling rate by free-free, free-bound and line emission. 
All the ions of  different elements in  different
ionization levels participate to the cooling  process. 
O$^+$ and O$^{++}$ ions peak at different temperatures.
Therefore, the electron temperature obtained from the [OIII]5007+4959/[OIII]4363 line ratios 
(Osterbrock 1974) is a valid approximation for the calculation of O/H from the [OIII]/\Hb line ratios, 
less precisely  for [OII]/\Hb.
O$^+$ and N$^+$ are linked by charge exchange reaction with H$^+$ because they correspond to a  similar first
ionization potential. Therefore O$^+$/O and N$^+$/N fractional abundances peak at the same
temperatures.

\begin{figure}
\centering
\includegraphics[width=9.6cm]{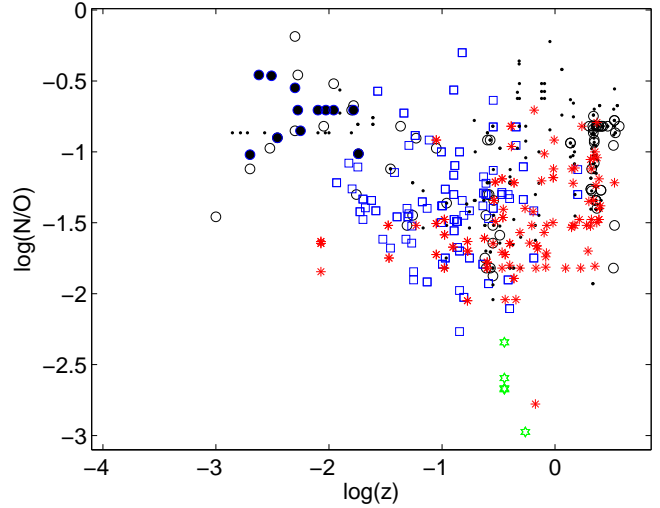}
\caption{Distribution of N/O as function of the redshift for SN host galaxies (squares),
long GRB hosts (asterisks), short GRB hosts (stars)
and  SB (dots), AGN (open circles) and LINERs  (filled circles) both local and at higher z.
}
\end{figure}

The N/O results calculated  until now (Contini 2017, 2016a, 2015 and references therein) are  summarised in Fig. 1
(top diagram),
where  the distribution of the N/O abundance ratios calculated for different types of host galaxies are shown as 
function of the redshift. 
We  distinguish   grossly  four main  types: 
SN hosts, LGRB hosts  and  galaxies  which were classified as SB  or as AGN + LINERs
according to  their line intensity ratios.
We would like to investigate   whether the N/O  wavy trend in Fig. 1 
is  determined by the behaviour of O/H, of N/H or  else, as function of z. 
Although some interesting features  appear in Fig. 1,  to  undertake a more detailed examination
of the N/O distribution 
on z, in this paper we extend our  investigation to  galaxies at lower z by  adding other 
samples  e.g. 
those  relative to HII  complexes in  "nearby" galaxies presented by Marino et al (2013) and  
in numerous low-luminosity  spirals and dwarf irregular galaxies at local distances observed
 by Berg et al (2012). Their infrared (IR)  luminosity ranges from -13 to -21.7 mag.
We  also added selected  AGN spectra at z$\leq$0.1 from Koski (1978, for Seyfert 2 galaxies), Cohen (1983), Kraemer et al (1994) and
Dopita et al (2015)  and the spectra of TypeIa SN hosts at z$\leq$0.03  reported by Gallagher et al (2005). 
Moreover, we  will  explore  star evolution in the host galaxies, considering the N/O   distribution
 as  function of metallicity (in term of the O/H relative abundance) on the basis
of Henry et al (2000),   MC02, etc. theories. MC02
 claim that at a given metallicity the N/O ratio shows a large dispersion at
 low as well as at high metallicity, not only due to the abundance determination, 
but principally to galaxy evolution models. We will compare our results with their models
in order to obtain some information about the star characteristics   throughout SN and GRB hosts.

In Sect. 2 we  briefly present  the  spectral line calculation method and
the  results of  modelling the samples of  HII region,  AGN observed line ratios  and TypeIa SN host spectra.
In Sect. 3 we examine  the distribution of N/O  versus z  considering previous and new results.
In Sect. 4 the N/O  versus log(O/H)+12 trends are investigated
on the basis of the results of chemical evolution models  calculated to "constrain the mode and
the history of star formation in starburst galaxies" by  MC02.
Concluding remarks follow in Sect. 5.

\begin{table*}
\centering
\caption{Modelling line ratios to \Hb =1 from the CALIFA HII regions (Marino et al 2013)}
%\tiny{
\begin{tabular}{lcccccccccccc} \hline  \hline
\  ID               &   z      & [OII]3727+     & [OIII]4363    &[OIII]5007+  & [NII]6583 &F$_{H\beta}^1$\\  \hline
\ MCG-01-54-016-001 &0.0098    & 3.83           & 0.041          &3.11       & 0.12      &139.78 \\ 
\ mc1               &          & 3.94           & 0.024         & 3.2        & 0.14      & \\
\ NGC 3991-001      & 0.010    & 2.88           & 0.018         & 2.73       & 0.24      & 1859.3\\
\ mc2               &          & 2.8            & 0.017         & 2.85       & 0.25      & \\
\ NGC 3991-002      &          & 3.98           & 0.016         & 2.18       & 0.30      & 1019.1 \\
\ mc3               &          & 3.96           & 0.023         & 2.3        &0.35       &\\
\ NGC 3991-007      &          & 5.18           & 0.013         &1.77        &0.37       & 233.58\\
\ mc4               &          & 5.             &0.03           &1.87        &0.33       &\\
\ NGC 7489-002      & 0.0208   & 3.35           &0.043          &3.19        &0.26       &87.13\\
\ mc5               &          & 3.6            &0.025          &3.16        &0.26       &\\
\ NGC 7489-007     &          &3.87            &0.08           &3.87        &0.26       &80.43\\
\ mc6               &          &3.99            &0.043          &3.83        &0.36      &\\
\ UGC 00312-001     &0.01456   &3.51            &0.015          & 2.93       &0.23      &512.98\\
\ mc7               &          &3.6             &0.02           &2.91        &0.25      &\\
\ UGC 00312-004     &          &3.07            &0.016          &2.68        &0.25      &296.63\\
\ mc8               &          &3.18            & 0.019         &2.66        &0.24      &\\
\ UGC 00312-005     &          &3.97            &0.018          &2.38        &0.26      &155.21\\
\ mc9               &          &3.7             &0.021          &2.33        &0.25      &\\
\ UGC 00312-007     &          &3.65            &0.018          &2.82        &0.24      &155.22\\
\ mc10              &          &3.5             &0.021          &2.8         &0.25      &\\
\ UGC 00312-008     &          &3.0             &0.034          &3.12        &0.19      &59.91\\
\ mc11              &          &3.2             &0.022          &3.15        &0.2       &\\
\ UGC 00312-013     &          &4.33            &0.033          &1.92        &0.29      &39.53\\
\ mc12              &          &4.48            &0.031          &1.8         &0.26      &\\
\ UGC 8733-004      &  0.0078  &2.44            &0.021          &2.25        &0.26      &46.46\\
\ mc13              &          &2.5             &0.017          &2.32        & 0.25     &\\
\ UGC 10331-004     &0.0149    &5.78            &0.019          &2.12        &0.39      &98.87\\
\ mc14              &          &5.3             &0.023          &2.17        &0.4       &\\
\ UGC 10796-001     & 0.0103   &3.14            &0.032          &1.91        &0.32      &94.54\\
\ mc15              &          &3.3             &0.022          &1.87        & 0.3      &\\
\ UGC 12494-001     & 0.014    & 2.51           &0.064          &4.67        &0.12      &74.13 \\
\ mc16              &          &2.8             &0.041          &4.99        &0.124 \\ \hline
\end{tabular}

$^1$ in 10$^{-16}$ \erg

\end{table*}

\begin{table*}
\centering
\caption{Models used  to reproduce the spectra presented in Table 2}
%\tiny{
\begin{tabular}{lccccccccccccccc} \hline  \hline
\  model &  \Vs   &\n0    & $D$        & N/H       & O/H        & \Ts      & $U$ & \Hb $^1$  \\
\        & \kms   & \cm3  & 10$^{18}$cm& 10$^{-4}$ & 10$^{-4}$  & 10$^4$ K & -   & \erg \\ \hline
\  mc1   & 100    & 120   & 0.7        & 0.08      & 6.0        & 8.3      & 0.009& 0.008\\
\  mc2   & 100    & 120   & 0.6        & 0.2       & 6.6        &6.6       &0.016 & 0.011\\
\  mc3   & 100    & 120   & 0.6        & 0.2       &6.4         & 7.4      &0.0072&0.007 \\
\  mc4   & 100    & 100   &0.7         & 0.15      &6.2         & 8.2      &0.004 &0.0038 \\
\  mc5   & 100    & 60    & 0.6        & 0.2       &6.8         & 7.3      &0.01    &0.0035  \\
\  mc6$^2$&200    &100    &0.093       &0.2        &6.3         & 7.8      &0.015   &0.0137  \\
\  mc7   &100     &100    &1.0         &0.18       &6.2         & 7.4      &0.0075  &0.0062  \\
\  mc8   &100     &100    &1.5         &0.16       &6.3         &7.8       &0.0095  &0.0077 \\
\  mc9   &100     &100    &1.5         &0.15       &6.5         &8.1       &0.006   &0.0063 \\
\  mc10  &100     &100    &1.5         &0.16       &6.5         &8.2       &0.0085  &0.0072 \\
\  mc11  &90      & 70    &2.6         &0.14       & 6.2        &8.6       &0.0085  &0.0046 \\
\  mc12  &100     & 70    &2.2         &0.15       &6.9         &8.6       &0.0036  &0.003 \\
\  mc13  &100     & 70    & 5.4        &0.2        & 6.3        &8.8       &0.0082  &0.0063 \\
\  mc14  & 90     &170    & 1.8        &0.16       &6.6         &11.       &0.004   &0.0093\\
\  mc15  &100     & 67    &3.4         &0.2        &6.6         &8.3       & 0.005  &0.0039\\
\  mc16  &100     &70     &0.19        &0.12       &6.0         &7.5       &0.012   &0.0037\\ \hline
\end{tabular}

$^1$ calculated at the nebula ; $^2$ calculated by  inflow.

\end{table*}

\begin{table*}
\centering
\caption{Modelling  line ratios to \Hb =1 from the Berg et al (2012) sample}
%\tiny{
\begin{tabular}{lccccccccccccccc} \hline  \hline
\ ID        & z   & [OII]  & [NeIII] & \Hg & [OIII]&[OIII] & HeI & [SIII]& [OI] & [NII]& \Ha &[SII]&[SII]\\ 
\           &     & $^1$   & $^2$    &$^3$ &$^4$   &$^5$    &$^6$& $^7$  &$^8$  &$^9$  &$^{10}$& $^{11}$&$^{12}$ \\  \hline
\ NGC521A   &0.0022&1.59   &0.67     & 0.48& 0.09  &4.85    &0.08 & 0.03  &0.03  &0.05  &2.77 &0.11 &0.07 \\
\ mb1      &     & 1.67   &0.34     &0.46 &0.15   &4.76    &0.16 & 0.03  &0.001 &0.08  &3.   &0.07 &0.04 \\
\ UGC695E   &0.00209&3.25    &0.55     &0.53 &0.04   &2.16    &0.10 &0.     &0.1   &0.16   &2.86 &0.43 &0.27\\
\ mb2     &     &3.0     &0.43     &0.46 &0.02   &2.4     &0.25 &0.     &0.3   &0.17   &3.   &0.55 &0.46 \\  
\ UGB1056A &0.00198&3.32    &0.64     &0.49 &0.04   &3.13    &0.13 &0.     &0.09  &0.19   &3.   &0.4  &0.31 \\
\ mb3     &     &3.2     &0.77     &0.46 &0.022  &3.3     &0.2  &0.     &0.24  &0.18   &3.   &0.44 &0.4 \\
\ UGB1056B &0.00198&2.8     & 0.84    &0.50 &0.04   &4.36    &0.11 &0.02   &0.17  &0.16   &2.82 &0.26 &0.17\\
\ mb4     &     &2.71    &0.8      &0.46 &0.022  &4.3     &0.19 &0.     &0.2   &0.17   &2.96 &0.29 &0.27\\ 
\ UGC1176A  &0.0021&2.27    &0.87     &0.46 &0.05   &4.81    &0.097&0.019  &0.1   &0.154  &2.82 &0.17 &0.12\\
\ mb5     &     &2.5     &0.78     &0.46 &0.022  &4.76    &0.18 &0.0    &0.18  &0.155  &2.96 &0.14 &0.13\\  
\ UGC784A   &0.0007&2.19    &0.53     &0.46 &0.05   &5.5     &0.11 &0.02   &0.073 &0.145  &2.89 &0.2  &0.15\\
\ mb6     &     &2.3     &0.83     &0.46 &0.023  &5.54    &0.17 &0.0    &0.17  &0.14   &2.9  &0.19 &0.18\\       
\ NGC784B   &0.0007&2.61    &0.8      &0.48 &0.05   &4.44    &0.095&0.015  &0.047 &0.1    &2.81 &0.2  &0.15\\
\ mb7     &     &2.6     &0.75     &0.46 &0.02   &4.47    &0.18 &0.0    &0.19  &0.11   &2.96 &0.18 &0.17\\
\ UGC2716A  &0.00137   &2.08    &0.7      &0.51 &0.06   &5.69    &0.11 &0.022  &0.045 &0.123  &2.86 &0.2  &0.14 \\
\ mb8     &     &2.2     &0.8      &0.46 &0.023  &5.8     &0.14 &0.001  &0.1   &0.12   &2.9  &0.18 &0.16 \\
\ NGC2537A  &0.0014     & 3.42   &0.45     &0.47 &0.012  &2.85    &0.12 & 0.016 &0.058 &0.63  &2.87 &0.36 &0.27 \\
\ mb9     &     & 3.5    &0.43     &0.46 &0.036  &3.0     &0.15 &-      &0.08  &0.7   &3.   &0.37 &0.34 \\
\ NGC2537B  &0.0014     & 2.79   &0.26     &0.44 &0.016  &2.38    &0.11 &0.013  &0.025 &0.54  &2.79 &0.26 &0.18 \\
\ mb10     &     & 2.99   &0.36     &0.46 &0.03   &2.6     &0.16 &0.001 &0.06  &0.6   &2.98 &0.33 &0.3 \\
\ UGC4278B  &0.0018     &2.03    &0.6      &0.46 &0.033  &2.54    &0.08 &0.013  &0.038 &0.08  &2.8  &0.18 &0.13 \\
\ mb11     &     &1.83    &0.3      &0.46 &0.032  &2.67    &0.16 &0.01   &0.024 &0.09  &3.   &0.17 &0.16 \\
\ UGC4278A  &0.0018     &1.68    &0.43     &0.44 &0.046  &3.36    &0.1  &0.013  &0.031 &0.071 &2.86 &0.14 &0.10\\
\ mb12     &     &1.56    &0.34     &0.45 &0.032  &3.44    &0.16 &0.01   &0.02  &0.073 &3.   &0.14 &0.13 \\
\ NGC2552A  &0.00175     &2.39    &0.75     &0.47 &0.021  &3.88    &0.10 &0.015  &0.05  &0.32  &2.86 &0.34 &0.25\\
\ mb13     &     &2.2     &0.4      &0.46 &0.03   &3.8     &0.16 &0.019  &0.04  &0.32  &3.   &0.37 &0.34 \\
\ UGC4393B  &0.0071     &2.6     &0.66     &0.47 &0.034  &4.38    &0.12 &0.011  &0.07  &0.39  &2.86 &0.28 &0.19\\
\ mb14     &     &2.9     &0.54     &0.46 &0.04   &4.5     &0.158&0.01   &0.05  &0.4   &3.   &0.23 &0.21\\
\ UGC4393C  &0.0071     &4.06    &0.5      &0.48 &0.036  &3.45    &0.11 &0.017  &0.11  &0.4   &2.87 &0.39 &0.28\\
\ mb15     &     &4.0     &0.44     &0.46 &0.044  &3.68    &0.15 &0.014  &0.06  &0.5   &2.96 &0.26 &0.22\\
\ CGCG0035-007A&0.0019  &4.00    &-        &0.46 &0.06   &2.71    &0.079&-      &-     &0.27  &2.87 &0.49 &0.33\\
\ mb16     &     &4.2     &-        &0.46 &0.06   &2.68    &0.16 &-      &-     &0.27  &2.96 &0.5  &0.46 \\ 
\ UGC5139A  &0.00046     &1.88    &0.54     &0.48 &0.051  &4.9     & -   &0.02   &0.1   &0.09  &2.83 &0.14 &0.09 \\
\ mb17     &     &1.7     &0.44     &0.45 &0.049  &4.9     &0.16 &0.01   &0.021 &0.097 &3.   &0.14 &0.13 \\
\ IC559     &0.0018     &3.12    &0.54     &0.45 &0.028  &3.73    &0.09 &-      &-     &0.18  &2.86 &0.35 &0.23 \\
\ mb18     &     &2.9     &0.45     &0.46 &0.044  &3.7     &0.15 &-      &-     &0.17  &2.98 &0.3  &0.28 \\
\ NGC5272A &0.0017&1.06    &0.56     &0.46 &0.083  &6.60    &0.1  &0.02   &0.017 &0.045 &2.83 &0.08&0.05 \\
\ mb19     &     &1.3     &0.58     &0.45 &0.07   &6.6     &0.16 &0.014  &0.011 &0.06  &3.   &0.13 &0.12  \\
\ NGC5340A  &0.000017     &0.58    &0.61     &0.47 & 0.06  &2.53    &0.09 & -     &  -   &0.016 &2.81 &0.05 &0.05 \\
\ mb20    &     &0.54    &0.36     &0.46 &0.03   &2.62    &0.17 & -     &-     &0.015 & 3.1 &0.06 &0.05 \\
\ UGC5423A  &0.00116     &2.08    & 0.6     &0.46 &0.07   &4.65    &0.1  &0.02   &0.1   &0.15  &2.86 &0.31 &0.22  \\
\ mb21    &     &2.2     &0.5      &0.46 &0.044  &4.55    &0.15 &0.02   &0.03  &0.18  & 3.  &0.33 &0.32 \\
\ UGC5423B  &0.00116     &1.78    &0.78     &0.46 &0.07   &4.95    &0.11 &0.02   &0.03  &0.11  &2.86 &0.15 &0.11   \\
\ mb22    &     &1.79    &0.6      &0.46 &0.042  &5.0     &0.16 &0.02   &0.02  &0.14  &3.   &0.2  &0.19   \\
\ NGC5672A  &0.023     &3.28    &0.39     &0.42 &0.04   &3.36    &0.11 &-      &0.04  &0.35  &2.87 &0.31 &0.22  \\
\ mb23    &     &3.0     &0.39     &0.46 &0.04   &3.7     &0.16 &0.015  &0.035 &0.38  &2.98 &0.26 &0.22  \\
\ UGC5692A  &0.00019     &2.47    &0.21     &0.54 &0.06   &2.27    &0.14 &-      &-     &0.48  &2.79 &0.61 &0.4  \\
\ mb24    &     &2.37    &0.27     &0.46 &0.03   &2.36    &0.16 &-      &-     &0.43  &3.   &0.4  &0.39  \\
\ UGC5797A  &0.0024     &1.59    &0.84     &0.47 &0.09   &7.38    &0.09 &0.02   &0.1   &0.12  &2.84 &0.22 &0.15   \\
\ mb25    &     &1.5     &0.64     &0.45 &0.12   &7.45    &0.16 &0.025  &0.004 &0.1   &3.1  &0.11 &0.1    \\
\ UGC5923A  &0.00237     &4.06    &0.44     &0.51 &0.06   &3.32    &-    &0.02   &0.08  &0.31  &2.78 &0.31 &0.22   \\
\ mb26    &     &4.4     &0.54     &0.46 &0.052  &3.4     &0.16 &0.02   &0.18  &0.27  &2.96 &0.37 &0.31  \\ 
\ NGC3741A  &0.000764     &1.6     &0.73     &0.46 &0.06   &3.78    &0.1  &0.018  &0.026 &0.068 &2.83 &0.12 &0.09  \\
\ mb27  &     & 1.8    &0.6      &0.45 &0.031  &3.64    &0.16 &0.01   &0.021 &0.09  &3.1  &0.13 &0.12 \\
\ NGC3738A  &0.00076     &2.91    &0.67     &0.47 &0.03   &3.94    &0.12 &0.02   &0.07  &0.25  &2.83 &0.32 &0.23  \\
\ mb28 &     &3.1     &0.5      &0.46 &0.039  &3.82    &0.16 &0.013  &0.056 &0.2   &3.   &0.24 &0.22  \\
\ NGC3738B  &0.00076     &3.45    &0.55     &0.46 &0.04   &4.14    &0.15 &-      &0.08  &0.26  &2.82 &0.29 &0.21  \\
\ mb29 &     &3.5     &0.54     &0.46 &0.044  &4.29    &0.15 &0.014  &0.063 &0.3   &2.97 &0.27 &0.24 \\
\ UGC6817   &0.0008 &0.94    &0.78     &0.46 &0.068  &3.86    &0.1  &0.015  &0.019 &0.038 &2.83 &0.07&0.05\\
\ mb30 &     &0.98    &0.58     &0.45 &0.04   &3.9     &0.17 &0.007  &0.01  &0.06  &3.1  &0.08 &0.07\\ \hline

\end{tabular}

 $^1$ 3727+ ;  $^2$ 3869+ ; $^3$4340 ; $^4$ 4363 ;   $^5$ 5007+ ;    $^6$ 5876 ; $^7$ 6312 ;  $^8$ 6300+ ; 
$^9$ 6548+ ; $^{10}$ 6563 ;$^{11}$ 6717 ; $^{12}$ 6731 ;

\end{table*}

\begin{table*}
\centering
                   Table 3 - continued
\\
\begin{tabular}{lcccccccccccccccc} \hline  \hline
\ ID        & z   & [OII]  & [NeIII] & \Hg & [OIII]&[OIII] & HeI & [SIII]& [OI] & [NII]& \Ha &[SII]&[SII]\\ 
\           &     & $^1$   & $^2$    &$^3$ &$^4$   &$^5$    &$^6$& $^7$  &$^8$  &$^9$  &$^{10}$& $^{11}$&$^{12}$ \\  \hline
\ UGC6900A  &0.00197     &4.13    & -       &0.55 &-      &0.83    &-    &-      &0.3   &0.68  &2.83  &1.11&0.6 \\ 
\ mb31 &     &4.1     &-        &0.45 &-      &0.88    &-    &-      &0.5   &0.66  &3.    &1.0 &1.5  \\
\ NGC4163A  &0.00055     &3.72    &-        &0.31 &0.014  &0.67    &0.10 &-      &0.07  &0.02  &2.79  &-   &-    \\
\ mb32 &     &3.98    &-        &0.45 &0.05   &0.6     &0.014&0.003  &0.4   &0.2   &3.    &-   &-     \\ 
\ CGCG269049C&0.00053    &1.76    &-        &0.29 &0.05   &2.05    &-    &-      &-     &0.02  &2.79  &-   &-     \\
\ mb33 &     &1.5     &-        &0.45 &0.02   &2.0     &-    &-      &-     &0.02  &3.    &-   &-     \\
\ CGCG269049A&0.00055    &1.0     &0.64     &0.46 &0.062  &3.35    &0.097&0.012  &0.011 &0.043 &2.83  &0.06&0.05 \\
\ mb34 &     &0.8     &0.5      &0.45 &0.04   &3.58    &0.17 &0.005  &0.008 &0.04  &3.1   &0.06&0.05 \\
\ UGC7577A  &0.00065     &1.65    &0.81     &0.47 &0.08   &7.0     &0.11 & -     &-     &0.12  &2.79  &0.2  &0.15\\
\ mb35 &     &1.9     &0.98     &0.46 &0.063  &7.09    &0.15 &-      &-     &0.13  &2.98  &0.21 &0.21 \\
\ NGC4449C  &0.00069     &3.26    &0.38     &0.45 &0.02   &3.0     &0.12 &0.014  &0.095 &0.3   &2.84  &0.35 &0.25\\
\ mb36 &     &3.0     &0.39     &0.45 &0.036  &3.16    &0.16 &0.012  &0.044 &0.2   &2.99  &0.22 &0.2 \\
\ NGC4449B  &0.00069     &3.0     &0.36     &0.48 &0.02   &3.38    &0.106&0.012  &0.065 &0.28  &2.86  &0.28 &0.2 \\
\ mb37 &     &3.0     &0.39     &0.46 &0.036  &3.25    &0.16 &0.012  &0.05  &0.25  &2.99  &0.22 &0.2  \\  
\ NGC449A   &0.00069     &2.4     &0.42     &0.47 &0.019  &4.61    &0.11 &0.014  &0.037 &0.211 &2.83  &0.18 &0.13 \\
\ mb38 &     &2.7     &0.49     &0.46 &0.04   &4.4     &0.157&0.013  &0.22  &0.22  &2.99  &0.21 &0.18\\
\ UGC7605   &0.001     & 1.91   &0.36     &0.46 &0.04   &3.09    &0.07 &0.01   &0.05  &0.081 &2.83  &0.15 &0.09\\
\ mb39 &     &1.8     &0.3      &0.45 &0.033  &3.03    &0.16 &0.01   &0.02  &0.1   &3.    &0.19 &0.16 \\
\ UGC7639A  &0.00127     &3.98    &-        &0.44 &0.05   &2.81    &-    &-      &0.09  &0.28  &2.83  &0.45 &0.28\\
\ mb40 &     & 4.0    &-        &0.46 &0.05   &2.5     &-    &-      &0.1   &0.25  &2.98  &0.5  &0.4 \\ 
\ NGC4656A  &0.00216     &0.8     &0.79     &0.45 &0.09   &9.      &-    &0.03   &0.03  &0.032 &2.86  &0.07 &0.04 \\
\ mb41 &     &1.6     &0.8      &0.45 &0.1    &8.9     &-    &0.01   &0.02  &0.06  &3.    &0.1  &0.1 \\
\ UGC8201A  &0.0001     &1.62    &0.44     &0.47 &0.05   &3.93    &0.11 &0.02   &0.1   &0.04  &2.81  &0.1  &0.07\\
\ mb42 &     &1.75    &0.42     &0.45 &0.033  &3.89    &0.16 &0.01   &0.016 &0.04  &3.1   &0.1  &0.1 \\
\ UGC8245A  &0.00023     &2.81    &-        &0.41 &0.03   &1.68    &0.10 &-      &0.23  &0.16  &2.83  &0.27 &0.18 \\ 
\ mb43 &     &2.73    &-        &0.45 &0.05   &1.77    &0.16 &-      &0.08  &0.18  &3.    &0.04 &0.37 \\
\ UGC8508A  &0.000207     &1.47    &0.78     &0.45 &0.06   &4.33    &0.10 &0.02   &0.06  &0.05  &2.83  &0.12 &0.08 \\
\ mb44 &     &1.47    &0.5      &0.45 &0.034  &4.3     &0.16 &0.01   &0.014 &0.05  &3.    &0.11 &0.1 \\ 
\ UGC8638A  &0.000915     &1.77    &0.9      &0.45 &0.055  &5.56    &0.10 &0.02   &0.06  &0.05  &2.83  &0.12 &0.08 \\
\ mb45 &     &1.9     &0.8      &0.45 &0.05   &5.7     &0.1  &0.01   &0.02  &0.04  &3.    &0.13 &0.12 \\
\ UGC8638B  &0.000915     &1.75    &0.34     &0.47 &0.06   &5.54    &0.11 &0.018  &0.017 &0.08  &2.81  &0.11 &0.09 \\
\ mb46 &     &1.75    &0.5      &0.45 &0.053  &5.8     &0.016&0.012  &0.017 &0.09  &3.    &0.12 &0.11 \\
\ UGC8837A  &0.00048     &3.46    &0.23     &0.47 &0.019  &1.68    &0.091&0.018  &0.064 &0.269 &2.83  &0.35 &0.25\\
\ mb47 &     &3.3     &0.35     &0.45 &0.06   &1.66    &0.17 &0.01   &0.128 &0.21  &3.    &0.37 &0.32\\       
\ NGC5477A  &0.001     &1.31    &0.66     &0.47 &0.06   &6.19    &0.11 &0.02   &0.064 &0.062 &2.84  &0.08&0.06\\
\ mb48 &     &1.23    &0.58     &0.45 &0.07   &6.37    &0.158&0.01   &0.01  &0.08  &3.    &0.1  &0.1\\
\ UGC9405A  &0.00074     &3.66    &-        &0.35 &-      &1.9     &-    &-      &-     &-     &2.79  &0.79 &0.55\\
\ mb49 &     &3.6     &-        &0.46 &-      &2.2     &-    &-      &-     &-     &2.98  &0.5  &0.45 \\
\ UGC10818A &0.03     &2.73    &0.48     &0.45 &0.02   &2.96    &0.10 &0.01   &0.08  &0.33  &2.79  &-    &-    \\
\ mb50 &     &2.8     &0.46     &0.46 &0.04   &3.91    &0.16 &0.01   &0.05  &0.27  &3.    &-    &-    \\
\ KKH098A   &0.0004     &1.85    &0.3      &0.58 &-      &2.54    &0.08 &-      &-     &0.12  &2.79  &0.15 &0.10 \\
\ mb51 &     &1.64    &0.29     &0.45 &0.028  &2.6     &0.16 &-      &-     &0.16  &3.    &0.15 &0.13 \\ \hline  
\end{tabular}

$^1$ 3727+ ;  $^2$ 3869+ ; $^3$4340 ; $^4$ 4363 ;   $^5$ 5007+ ;    $^6$ 5876 ; $^7$ 6312 ;  $^8$ 6300+ ; 
$^9$ 6548+ ; $^{10}$ 6563 ;$^{11}$ 6717 ; $^{12}$ 6731 ;

\end{table*}

\begin{table*}
\centering
\caption{Models used  to reproduce the  spectra presented in Table 3}
%\tiny{
\begin{tabular}{lccccccccccccccc} \hline  \hline
\  model &  \Vs   &\n0    & $D$        & N/H       & O/H       &Ne/H      &  S/H      & \Ts      & $U$  & \Hb $^1$ & \\
\        & \kms   & \cm3  & 10$^{18}$cm& 10$^{-4}$ & 10$^{-4}$ &10$^{-4}$ &10$^{-4}$  & 10$^4$ K & -    & \erg &  \\ \hline
\  mb1  & 40     &100    &0.05        & 0.2       &6.6        &1.        &0.23       &4.4       &0.07  &4.e-4 \\
\ mb2 &100     &60     &7.          &0.08       &6.         &1.        &0.05       &8.6       &0.005  &0.0045\\
\ mb3 &100     &80     &2.          &0.06       &4.         &1.        &0.03       &8.7       &0.007 &0.005\\
\  mb4 &100     &80     &2.          &0.07       &4.         &1.        &0.02       &8.3       &0.012 &0.007\\
\ mb5 & 100    &80     &2.          &0.07       &4.2        &1.        &0.01       &8.        &0.016 &0.0077\\
\  mb6 &100     &80     &2.          &0.07       &4.2        &1.        &0.013      &8.        &0.024 &0.009\\   
\  mb7 &100     &80     &2.          &0.05       &4.2        &1.        &0.013      &8.        &0.014 &0.0073\\
\  mb8 & 100    &70     &0.4         &0.07       &4.2        &1.        &0.013      &7.        &0.03  &0.007\\
\  mb9 &130     &60     &0.3         &0.25       &4.6        &0.7       &0.1        &8.6       &0.005 &0.0035 \\
\  mb10 &130     &60     &0.35        &0.25       &4.6        &0.7       &0.1        &8.6       &0.005 &0.004 \\
\  mb11 &130     &60     &0.58        &0.07       &5.3        &1.        &0.1        &8.6       &0.007 &0.0063 \\
\  mb12 &130     & 60    &0.71        &0.07       & 6.6       &1.        &0.1        &8.9       &0.01  &0.01 \\
\  mb13 &130     &60     &0.51        &0.22       & 6.8       &1.        &0.2        &8.9       &0.0086&0.007\\
\  mb14 &130     &60     &0.53        &0.22       &6.3        &1.        &0.1        &9.1       &0.0086&0.0068\\
\ mb15 &100     &60     &0.42        &0.22       &6.6        &1.        &0.1        &6.6       &0.005 &0.003\\
\  mb16 &100     &60     &0.3         &0.1        &6.6        &1.        &0.15       &7.6       &0.003 &0.002 \\
\  mb17 &130     &60     &0.82        &0.09       &6.8        &1.        &0.1        &9.3       &0.014 &0.012\\
\  mb18 &130     &60     &0.42        &0.09       &6.8        &1.        &0.13       &9.3       &0.006 &0.0045 \\
\  mb19&120     &60     &1.16        &0.07       &6.5        &1.        &0.13       &8.3       &0.025 &0.015   \\
\  mb20& 100    &60     &2.14        &0.07       &6.6        &1.6       &0.13       &8.9       &0.016 &0.02  \\
\  mb21& 130    &70     &0.39        &0.12       &6.5        &1.        &0.2        &8.4       &0.0096&0.0067 \\
\ mb22& 130    &70     &0.54        &0.12       &6.7        &1.        &0.15       &8.4       &0.015 &0.012  \\
\ mb23& 100    &50     &0.69        &0.22       &6.6        &1.        &0.13       &6.7       &0.006 &0.003  \\
\ mb24& 100    &80     &0.32        &0.28       &6.6        &1.        &0.23       &6.7       &0.0058&0.0058 \\
\ mb25& 100    &65     &1.45        &0.13       &6.2        &1.        &0.25       &5.2       &0.032 &0.017  \\
\ mb26& 100    &60     &0.37        &0.1        &6.3        &1.        &0.1        &8.        &0.0036&0.0024 \\
\ mb27&100    &70     &0.74        &0.08       &6.6        &1.7       &0.1        &8.        &0.009 &0.009  \\
\ mb28& 100    &70     &0.47        &0.1        &6.3        &1.        &0.1        &8.        &0.0058&0.0052 \\
\ mb29& 100    &70     &0.41        &0.15       &6.3        &1.        &0.1        &8.        &0.0058&0.0046\\ 
\ mb30& 100    &50     &1.85        &0.1        &6.6        &1.7       &0.1        &8.9       &0.012 &0.0092\\
\ mb31& 120    &350    &0.091       &0.1        &3.8        &1.        &0.06       &0.7       &4.e-4 &0.0056 \\
\ mb32& 120    &350    &0.084       &0.1        &4.         &1.        &0.06       &0.7       &5.e-4 &0.0085 \\  
\ mb33& 100    &70     &0.65        &0.07       &6.6        &1.0       &0.13       &6.8       &0.007 &0.0086 \\
\ mb34& 100    &70     &1.75        &0.07       &6.6        &1.6       &0.08       &9.3       &0.014 &0.014 \\
\ mb35& 130    &80     &0.44        &0.1        &6.4        &1.5       &0.15       &8.7       &0.019 &0.013\\
\ mb36& 100    &70     &0.412       &0.1        &6.3        &1.        &0.1        &7.0       &0.0058&0.0049 \\
\ mb37& 100    &70     &0.411       &0.13       &6.4        &1.        &0.1        &7.0       &0.0058&0.0048\\
\ mb38& 100    &70     &0.67        &0.13       &6.4        &1.        &0.1        &8.0       &0.0069&0.0049\\
\ mb39& 100    &60     &0.76        &0.09       &6.6        &1.        &0.14       &7.6       &0.0063&0.003 \\
\ mb40& 100    &60     &0.344       &0.1        &6.6        &1.        &0.15       &5.6       &0.003 &0.0021\\  
\ mb41& 130    &80     &0.76        &0.06       &6.4        &1.        &0.08       &8.6       &0.04  &0.025 \\
\ mb42&120     &50     &1.          &0.09       &6.8        &1.2       &0.1        &7.1       &0.012 &0.0078\\
\ mb43&100     &70     &0.29        &0.1        &6.6        &0.8       &0.15       &8.2       &0.0026&0.0028\\
\ mb44&120     &60     &0.88        &0.09       &6.8        &1.2       &0.1        &8.        &0.014 &0.011\\
\ mb45&120     &60     &0.73        &0.09       &6.8        &1.6       &0.1        &8.        &0.014 &0.008\\
\ mb46&100     &60     &1.15        &0.09       &6.6        &1.        &0.1        &8.        &0.014 &0.0093\\
\ mb47&100     &60     &0.3         &0.1        &6.6        &1.        &0.1        &9.        &0.0016&0.0016\\
\ mb48&100     &70     &1.2         &0.1        &6.4        &1.        &0.1        &8.5       &0.022 &0.015 \\
\ mb49&100     &70     &0.23        &0.1        &6.6        &0.8       &0.15       &8.2       &0.0023&0.0018\\
\ mb50&90      &70     &0.46        &0.15       &6.6        &1.        &0.1        &8.4       &0.004 &0.0036 \\
\ mb51&90      &70     &0.85        &0.15       &6.6        &1.        &0.1        &9.9       &0.005 &0.0073 \\ \hline  

\end{tabular}

$^1$ calculated at the nebula

\end{table*}

\begin{table*}
\centering
\caption{Modelling line ratios to \Hb =1 of  AGN spectra at z$\leq$0.1 (Dors et al 2017)}
%\tiny{
\begin{tabular}{lcccccccccccc} \hline  \hline
\  ID    &   z  & [OII]3727+  &[OIII]5007+ & [NII]6583+& [SII] 6719+ &Ref \\  \hline
\ IZW92  &0.038 & 2.63        &13.5        &1.3        &0.77         & 1   \\
\  ms1   &      & 2.76        &13.7        &1.2        &0.97         &     \\
\ Mrk176 & 0.027&3.54         &18.15       &3.99       &1.10         &2    \\
\  ms2   &      &3.9          &18.0        &3.99       &1.1          &     \\
\ 3c33   & 0.0597&4.93        &16.78       &2.36       &1.6          &2    \\
\ ms3    &      &5.1          &16.87       &2.34       &1.5          &    \\
\ Mrk3   &0.0135 &3.52        &16.66       &4.19       &1.55         &2 \\
\ ms4    &      &3.4          &17.0        &4.2        &1.7          &  \\
\ Mrk573 &0.0172 &2.92        &16.0        &3.29       &1.55         &2  \\
\ ms5    &      &2.99         &16.2        &3.3        &1.58         &   \\
\ Mrk78  &0.037 &4.96         &15.82       &3.05       &1.62         &2\\
\ ms6    &      &4.74         &16.         &2.9        &1.7          & \\
\ Mrk34  & 0.051 &3.43        &15.14       &2.88       &1.62         &2\\
\ ms7    &      &3.7          &14.8        &2.9        &1.5          &  \\
\ Mrk1   &0.016 &2.78         &14.51       &2.98       &1.0          &2 \\
\ ms8    &      &2.97         &14.66       &3.97       &1.3          & \\
\ 3c433  &0.1   &6.17         &12.81       &7.22       &2.71         &2\\
\ ms9    &      &5.5          &12.9        &6.9        &2.9          & \\
\ Mrk270 &0.01  &5.44         &11.67       &3.88       &2.6          &2\\ 
\ ms10   &      &5.8          &11.65       &3.8        &2.6          & \\
\ 3c452  & 0.081&4.81         &9.25        &4.96       &1.87         &2\\
\ ms11   &      &4.83         &9.27        &4.9        &1.6          &  \\
\ Mrk198 & 0.024&2.51         &7.35        &2.95       &1.57         &2\\
\ ms12   &      &2.7          &7.1         &2.98       &1.56         & \\
\ Mrk268 &0.04  &3.75         &6.37        &6.41       &2.36         &2\\
\ ms13   &      &3.66         &6.22        &6.35       &2.3          & \\
\ NGC3227&0.00385&3.32        &14.3        &6.68       &2.5          &3\\
\ ms14   &      &3.1          &14.7        &6.8        &2.5          &  \\
\ Mrk6   &0.01  &2.45         &13.5        &2.38       &1.25         &3 \\
\ ms15   &      &2.6          &13.5        &2.34       &1.2          & \\
\ NGC7590&0.0052 &3.32        &4.63        &2.47       &1.77         &3\\
\ ms16   &      &3.7          &4.8         &2.7        &1.8          & \\
\ NGC613 &0.0049&2.42         &0.68        &2.27       &0.91         &4\\
\ ms17   &      &2.2          &0.7         &2.39       &0.8          & \\
\ IC1657 &0.0119&3.67         &2.38        &2.52       &1.85         &4\\
\ ms18   &      &3.4          &2.27        &2.59       &1.6          &  \\
\ IRAS1475&0.0177&1.78        &6.65        &2.25       &0.54         & 4\\
\ ms19   &      &1.9          &6.68        &2.3        &0.65         &  \\
\ IC1816 &0.0169&1.99         &13.7        &6.0        &2.16         &4 \\
\ ms20   &      &2.3          &13.6        &6.4        &2.3          & \\
\ NGC1125&0.0109&3.14         &8.21        &2.87       &1.38         &4 \\
\ ms21   &      &3.26         &8.5         &2.81       &1.48         & \\
\ MCG-06 &0.0152&2.92         &9.53        &3.47       &1.93         & 4 \\
\ ms22   &      &3.1          &9.34        &3.4        &1.9          &  \\
\ IRAS11 &0.014 &2.46         &10.49       &2.16       &1.34         & 4 \\
\ ms23   &      &3.08         &10.6        &2.2        &1.6          &  \\
\ ESO137 &0.0091&3.10         &12.4        &4.97       &2.36         & 4\\
\ ms24   &      &3.0          &12.47       &4.9        &2.3          &  \\
\ NGC6300&0.0037&5.41         &19.73       &6.62       &1.7          &4 \\
\ ms25   &      &5.0          &19.8        &6.3        &1.7          &  \\
\ ESO103 &0.013 &3.19         &9.8         &4.23       &2.06         & 4\\
\ ms26   &      &3.1          &10.         &4.3        &2.           & \\
\ NGC6926&0.0196&5.84         &5.84        &4.8        &2.5          & 4\\
\ ms27   &      &5.4          &6.          &4.87       &2.5          & \\
\ IC1368 &0.013 &4.56         &4.8         &4.51       &1.31         &4\\
\ ms28   &      &4.5          &4.8         &4.4        &1.24         & \\ \hline

\end{tabular}

 Ref : 1  Kraemer et al (1994) ; 2  Koski (1978); 3  Cohen (1983) ; 4  Dopita et al (2015)

\end{table*}

\begin{table*}
\centering
\caption{Models  used  to reproduce the spectra presented in Table 5}
%\tiny{
\begin{tabular}{lcccccccccccc} \hline  \hline
\ model& \Vs & \n0  & $F$  &$D$         & N/H        & O/H       & S/H       & \Hb (calc)$^1$ \\
\      & \kms& \cm3& $^2$ &10$^{17}$cm & 10$^{-4}$  & 10$^{-4}$ & 10$^{-4}$ & \erg   \\  \hline
\  ms1 &320    &330   &16.  &7.         &0.3         &5.8        &0.07       &0.25   \\ 
\  ms2 &120    &420   &10.  &0.5        &1.          &5.7        &0.036      &0.1    \\ 
\ ms3  &280    &200   &10.  &70.        &0.45        &6.6        &0.04       &0.12   \\
\ ms4  & 190   &470   &37.  &0.8        &1.          &6.6        &0.1        &0.33    \\         
\ ms5  &210    &470   &37.  &0.8        &0.8         &6.5        &0.1        &0.37   \\
\ ms6  &300    &200   &9.   &80.        &0.6         &6.6        &0.05       &0.12   \\
\ ms7  &280    &290   &12.  &20.        &0.6         &6.0        &0.05       &0.166   \\  
\ ms8  &320    &280   &16.  &0.7        &0.7         &5.3        &0.1        &0.2     \\
\ ms9  &210    &150   &1.8  &1.         &1.8         &7.0        &0.15       &0.029   \\
\ ms10 &210    &150   &1.5  &1.         &0.9         &6.6        &0.12       &0.026  \\
\ ms11 &330    &180   &2.4  &1.         &1.1         &6.6        &0.08       &0.05   \\
\ ms12 &400    &200   &4.4  &1.6        &1.          &6.6        &0.12       &0.12   \\
\ ms13 &380    &160   &2.   &1.2        &1.8         &6.6        &0.13       &0.053  \\ 
\ ms14 &310    &280   &13.  &0.7        &1.8         &6.6        &0.19       &0.19   \\
\ ms15 &330    &330   &17.  &0.7        &0.6         &6.0        &0.09       &0.27   \\
\ ms16 &380    &190   &1.9  &1.0        &0.7         &6.2        &0.12       &0.055  \\
\ ms17 &330    &330   &0.6  &0.7        &0.8         &6.0        &0.09       &0.0295 \\
\ ms18 &320    &320   &1.5  &0.7        &0.6         &6.6        &0.012      &0.048  \\
\ ms19 &360    &360   &8.4  &0.7        &0.8         &6.6        &0.07       &0.26   \\
\ ms20 &320    &370   &21.  &0.7        &1.5         &5.4        &0.17       &0.29   \\
\ ms21 &400    &180   &4.4  &1.6        &0.9         &6.7        &0.1        &0.11   \\
\ ms22 &400    &170   &4.8  &1.6        &1.1         &6.7        &0.13       &0.11   \\
\ ms23 &400    &190   &7.0  &1.6        &0.7         &6.8        &0.1        &0.155 \\ 
\ ms24 &300    &320   &10.  &0.8        &1.3         &6.8        &0.18       &0.18  \\
\ ms25 &100    &330   &4.6  &0.4        &2.0         &7.0        &0.15       &0.053  \\
\ ms26 &400    &170   &5.5  &1.6        &1.4         &6.8        &0.14       &0.12   \\
\ ms27 &380    &120   &0.98 &1.0        &1.1         &6.6        &0.13       &0.024  \\
\ ms28 &380    &160   &1.3  &1.0        &1.0         &6.2        &0.07       &0.035  \\ \hline

\end{tabular}

$^1$ calculated at the nebula; $^2$ in 10$^{10}$ photons cm$^{-2}$ s $^{-1}$ eV$^{-1}$ at the Lyman limit;

\end{table*}
   
\begin{table*}
\centering
\caption{Modelling line ratios to \Hb =1 of  TypeI SN host spectra at z$\leq$0.03 (Gallagher et al 2005)}
%\tiny{
\begin{tabular}{lcccccccccccc} \hline  \hline
\ galaxy   &   SN  &  z   &  [OII] &  [OIII] &  [OI] &  [NII] & \Ha &  [SII] &  [SII]  \\                
\          &       &      &3727+ &  5007+  &  6300 &  6548+ & 6563&   6717 &  6731   \\
\ NGC4536  & 1981B &0.006 &10.7    &  1.0    & 0.16  & 1.7    & 3.  & 0.65   &0.47     \\
\ mg1      &       &      &10.0    & 1.6     & 0.5   & 1.3    &3.2  & 0.69   &0.54     \\
\ NGC3627  & 1989B &0.0024&1.28    & 0.28    & 0.08  & 1.8    &3.   & 0.49   &0.4      \\
\ mg2      &       &      &1.34    & 0.34    & 0.06  & 1.9    &3.   & 0.35   &0.36     \\
\ NGC4639  & 1990N &0.003 &16.2    & 1.3     &0.3    &0.2     &3.   &0.88    &0.5     \\
\ mg3      &       &      &14.0    & 2.1     &1.     &0.4     &3.   &0.98    &0.75    \\
\ NGC4527  & 1991T &0.0058&6.8     & 1.56    &0.     &1.8     &3.   &0.43    &0.42   \\
\ mg4      &       &      &6.7     &1.33     &0.3    &2.0     &3.   &0.5     &0.4     \\
\ IC4232   & 1991U &0.031 &0.24    &0.4      &0.09   &2.2     &3.   &0.42    &1.3    \\
\ mg5      &       &      &0.7     &0.2      &0.03   &1.3     &3.   &0.42    &0.6    \\
\ Anon     & 1992ag&0.000 &3.5     &1.9      &0.07   &1.6     &3.   &1.      &0.45   \\
\ mg6      &       &      &3.7     &1.9      &0.14   &1.6     &3.   &0.6     &0.5    \\
\ CGCG224  & 1994Q &0.0295&4.1     &0.5      &0.11   &1.      &3.   &0.7     &0.44   \\
\ mg7      &       &      &4.3     &0.88     &0.2    &1.1     &3.   &0.56    &0.55   \\
\ NGC4495  & 1994S &0.015 &5.2     &1.2      &0.11   &1.6     &3.   &0.54    &0.7    \\
\ mg8      &       &      &5.2     &1.2      &0.2    &1.9     &3.   &0.64    &0.43   \\
\ NGC3370  &1994ae &0.0043&5.43    &0.51     &0.     &1.3     &3.   &0.66    &0.42   \\
\ mg9      &       &      &5.4     &1.       &0.3    &1.3     &3.   &0.64    &0.63   \\
\ NGC2441  &1995E  &0.012 &4.66    &0.74     &0.03   &1.9     &3.   &0.82    &0.6   \\
\ mg10     &       &      &4.7     &0.9      &0.2    &2.      &3.   &0.65    &0.64  \\
\ IC1844   &1995ak &0.0227&5.53    &0.57     &0.08   &1.45    &3.   &0.73    &0.48  \\
\ mg11     &       &      &5.6     &1.       &0.2    &1.36    &3.   &0.67    &0.65   \\
\ NGC3021  &1995al &0.0051&3.2     &0.45     &0.1    &1.6     &3.   &0.47    &0.4   \\
\ mg12     &       &      &3.3     &0.7      &0.15   &1.65    &3.   &0.42    &0.42   \\
\ Anon     &1996C  &0.000 &7.6     &0.57     &0.48   &2.5     &3.   &1.1     &0.66   \\  
\ mg13     &       &      &7.5     &1.2      &0.9    &2.7     &3.   &1.0     &0.96   \\
\ Anon     &1996bl &0.000 &31.     &2.16     &0.34   &0.24    &3.   &0.71    &0.48   \\
\ mg14     &       &      &26.     &3.6      &1.2    &0.38    &3.6  &0.8     &0.6    \\
\ NGC0673  &1996bo &0.0174&3.7     &0.92     &0.06   &1.46    &3.   &0.59    &0.49   \\
\ mg15     &       &      &3.8     &0.8      &0.2    &1.6     &3.   &0.49    &0.5    \\
\ UGC03432 &1996bv &0.0167&9.87    &2.1      &0.18   &1.05    &3.   &0.96    &0.84   \\
\ mg16     &       &      &9.9     &2.1      &0.7    &1.07    &3.   &0.98    &0.86   \\
\ NGC4680  &1997bp &0.0083&3.      &0.4      &0.3    &1.66    &3.   &0.56    &0.42   \\
\ mg17     &       &      &3.1     &0.68     &0.12   &1.56    &3.   &0.53    &0.49   \\
\ Anon     &1997br &0.000 &4.55    &2.96     &0.21   &1.33    &3.   &0.51    &0.48   \\
\ mg18     &       &      &4.57    &2.9      &0.2    &1.4     &3.   &0.53    &0.5    \\
\ NGC0105  &1997cw &0.0176&5.6     &1.98     &0.1    &1.7     &3.   &0.59    &0.45   \\
\ mg19     &       &      &5.3     &1.97     &0.3    &1.6     &3.   &0.54    &0.5    \\
\ UGC03845 & 1997do&0.01  &6.69    &1.4      &0.076  &0.98    &3.   &0.74    &0.57   \\
\ mg20     &       &      &6.5     &1.32     &0.4    &1.      &3.   &0.75    &0.65   \\
\ NGC6627  &1998V  &0.0176&3.44    &1.55     &0.096  &2.5     &3.   &0.55    &0.42   \\
\ mg21     &       &      &3.5     &1.6      &0.15   &2.7     &3..  &0.55    &0.5    \\
\ NGC4704  &1998ab &0.027 &2.18    &0.34     &0.075  &2.93    &3.   &1.47    &0.4    \\
\ mg22     &       &      &2.1     &0.45     &0.06   &2.7     &3.1  &0.8     &0.7    \\
\ NGC3982  &1998aq &0.0037&3.3     &0.79     &0.07   &1.56    &3.   &0.59    &0.4    \\
\ mg23     &       &      &3.4     &0.86     &0.12   &1.56    &3.   &0.56    &0.5    \\
\ UGC00139 &1998dk &0.0132&3.8     &0.8      &0.15   &1.29    &3.   &0.73    &0.52   \\
\ mg24     &       &      &3.7     &1.       &0.15   &1.2     &3.   &0.7     &0.62   \\
\ UGCA017  &1998dm &0.0065&6.66    &2.26     &0.1    &0.75    &3.   &0.72    &0.47   \\
\ mg25     &       &      &6.5     &2.47     &0.2    &0.73    &3.   &0.71    &0.72   \\
\ NGC0632  &1998es &0.011 &2.72    &0.57     &0.15   &1.59    &3.   &0.6     &0.46   \\
\ mg26     &       &      &2.5     &0.53     &0.0    &1.53    &3.   &0.57    &0.51   \\
\ NGC6063  &1999ac &0.0095&4.17    &0.32     &0.13   &1.7     &3.   &0.68    &1.   \\
\ mg27     &       &      &4.      &0.7      &0.27   &1.6     &3.   &0.65    &0.7    \\ \hline                   
\end{tabular}

\end{table*}

\begin{table*}
\centering
\caption{Models  used  to reproduce the spectra presented in Table 7}
%\tiny{
\begin{tabular}{lcccccccccccc} \hline  \hline
\  model  & \Vs  &\n0   &   $D$       & N/H      &O/H       &   S/H   & \Ts     &  $U$     &   \Hb $^1$   \\
\         & \kms & \cm3 & 10$^{18}$cm & 10${-4}$ &10$^{-4}$ &10$^{-4}$& 10$^4$K &10$^{-3}$ &10$^{-3}$ \erg \\ \hline
\  mg1    & 70   & 60   & 1.9         &0.3       &6.6       & 0.04    &8.6      &0.5       & 0.37   \\
\  mg2    &100   &100   & 3.9         &1.4       &6.        &0.15     &4.6      &0.5       &3.3     \\
\  mg3    &80    &60    &1.2          &0.06      &6.2       &0.03     &20.      &0.6       &0.42    \\
\  mg4    &80    &60    &1.8          &0.6       &6.6       &0.04     &7.5      &0.8       &0.63 \\
\  mg5    &100   &250   &1.4          &1.3       &5.3       &0.2      &3.2      &4.8       &20.    \\
\  mg6    &70    &80    &0.8          &0.7       &6.9       &0.06     &6.       &4.        &2.4    \\
\  mg7    &100   &100   &0.8          &0.4       &6.4       &0.06     &6.       &2.        &2.5    \\
\  mg8    &100   &100   &0.8          &0.6       &6.7       &0.06     &7.       &2.        &2.5   \\
\  mg9    &100   &100   &0.8          &0.4       &6.6       &0.06     &7.       &1.3       &1.8    \\
\  mg10   &100   &100   &0.8          &0.7       &6.6       &0.07     &6.4      &1.3       &1.85   \\
\  mg11   &100   &100   &0.8          &0.4       &6.5       &0.06     &7.3      &1.3       &1.8    \\
\  mg12   &100   &100   &0.8          &0.8       &6.5       &0.065    &4.8      &1.3       &1.9    \\
\  mg13   &100   &130   &1.8          &0.6       &6.6       &0.06     &17.      &1.        &3.1  \\
\ mg14    &80    &50    &0.4          &0.03      &6.6       &0.02     &30.      &0.8       &0.29  \\
\ mg15    &100   &130   &1.8          &0.6       &6.6       &0.06     &7.6      &1.3       &4.1   \\
\ mg16    &100   &80    &1.           &0.2       &6.6       &0.05     &13.      &1.3       &1.3   \\
\ mg17    &100   &80    &1.           &0.86      &6.6       &0.09     &4.6      &1.3       &1.6   \\
\ mg18    &100   &80    &1.           &0.5       &6.6       &0.045    &7.5      &5.        &3.5  \\
\ mg19    &100   &80    &1.           &0.5       &6.6       &0.045    &7.5      &3.        &2.5  \\
\ mg20    &100   &70    &1.           &0.3       &6.6       &0.06     &8.2      &1.        &1.   \\
\ mg21    &100   &70    &1.           &1.2       &6.6       &0.06     &6.2      &4.5       &2.86  \\
\ mg22    &100   &70    &1.           &1.8       &6.6       &0.16     &4.2      &3.3       &2.3   \\
\ mg23    &100   &70    &1.           &0.8       &6.6       &0.07     &5.2      &3.        &2.2   \\
\ mg24    &100   &70    &1.           &0.5       &6.6       &0.08     &5.5      &3.        &2.2   \\
\ mg25    &100   &70    &1.           &0.2       &6.6       &0.05     &8.5      &2.6       &1.9   \\
\ mg26    &100   &70    &1.           &0.9       &6.6       &0.1      &4.2      &2.6       &2.    \\
\ mg27    &100   &120   &1.           &0.6       &6.6       &0.08     &5.9      &1.3       &2.5   \\ \hline     

\end{tabular}

$^1$ calculated at the nebula

\end{table*}

\begin{figure*}
\centering
\includegraphics[width=8.8cm]{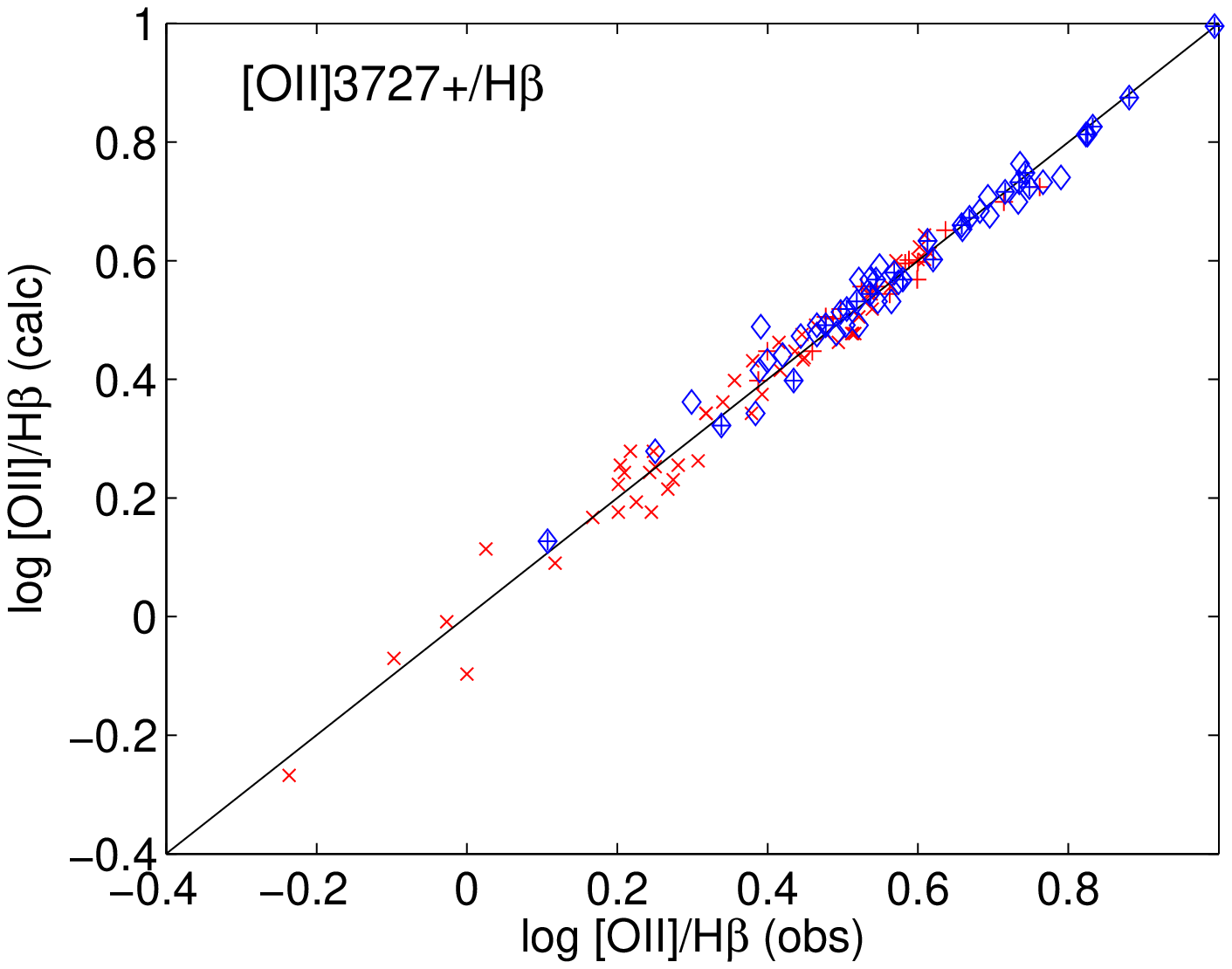}
\includegraphics[width=8.8cm]{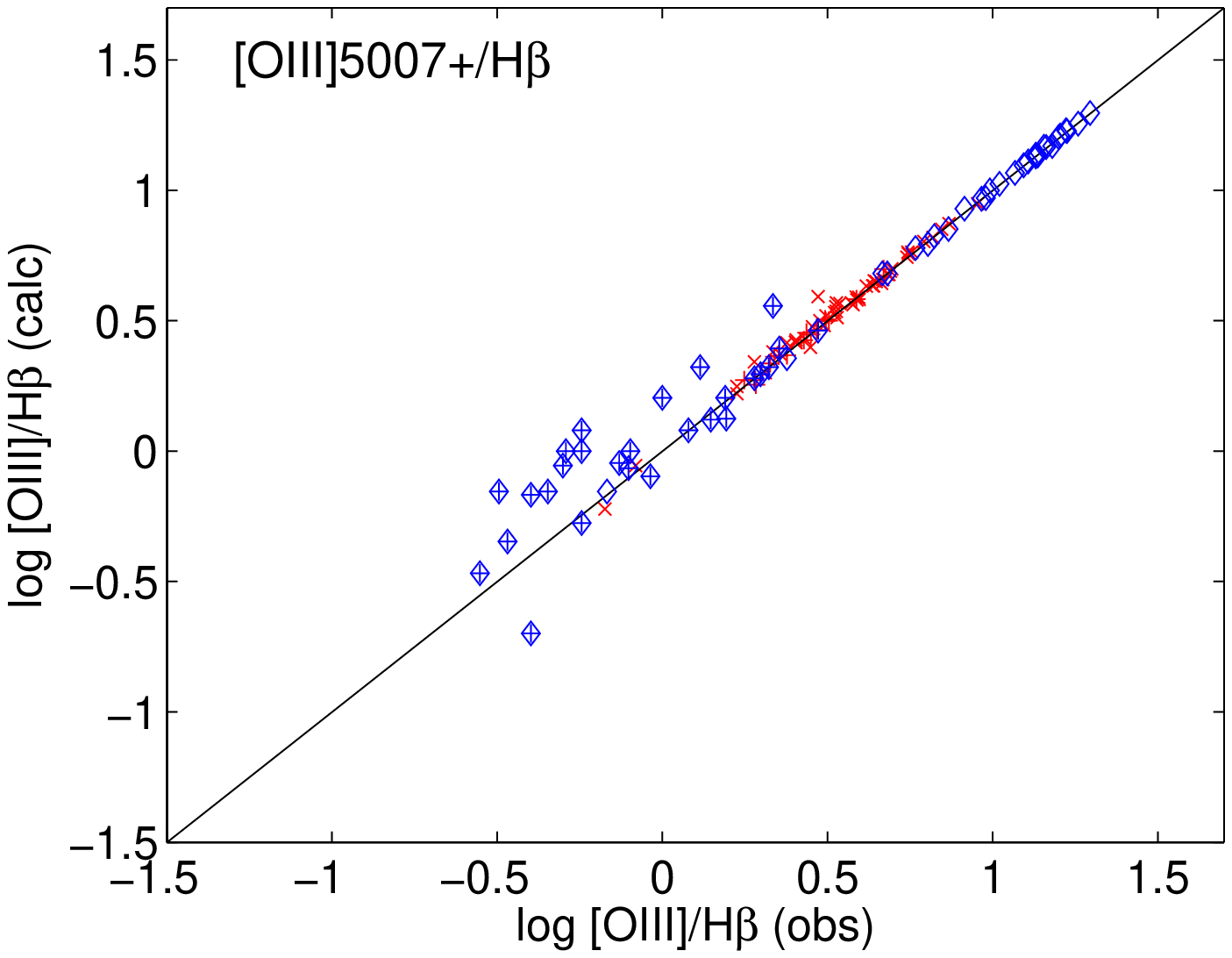}
\includegraphics[width=8.8cm]{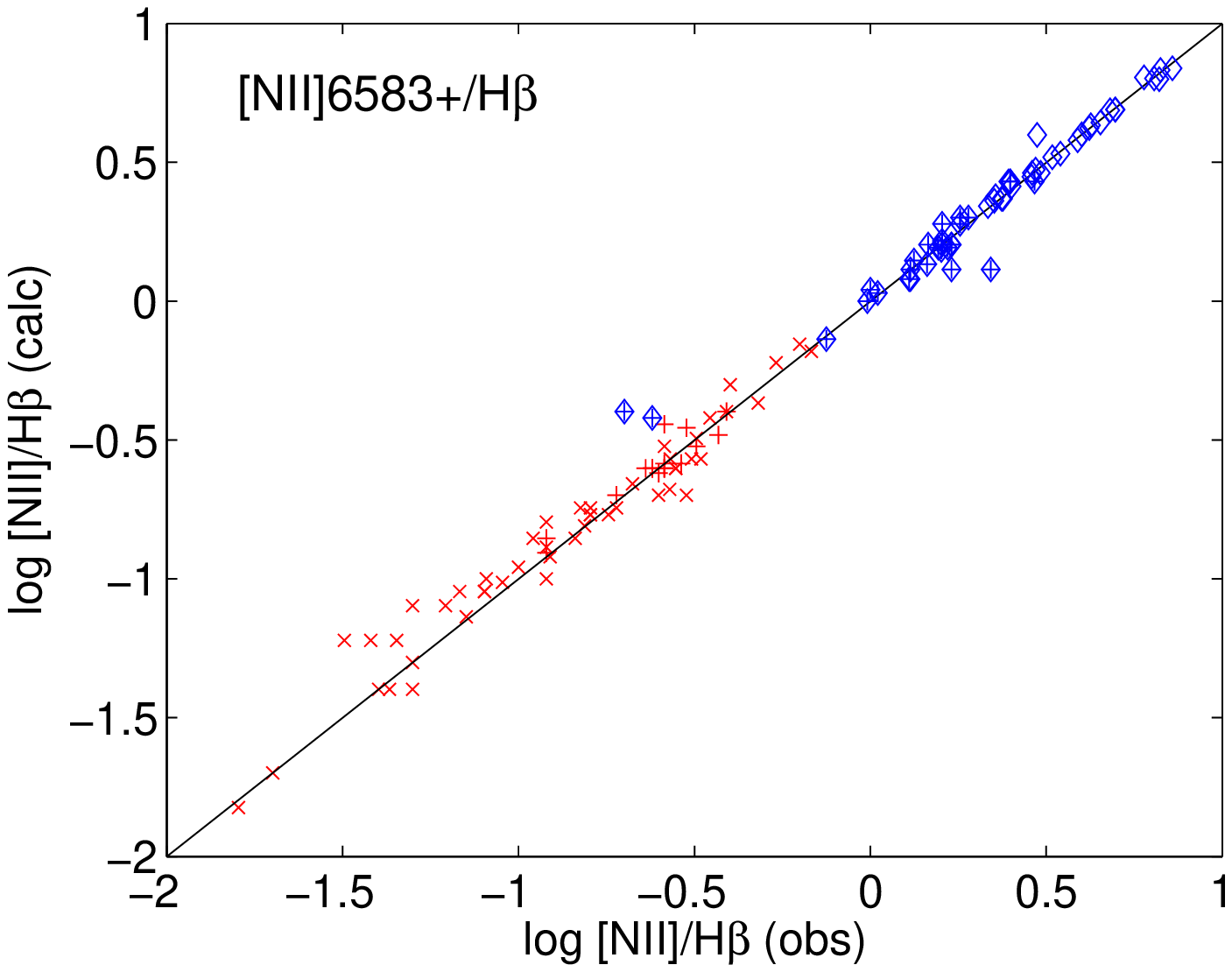}
\includegraphics[width=8.8cm]{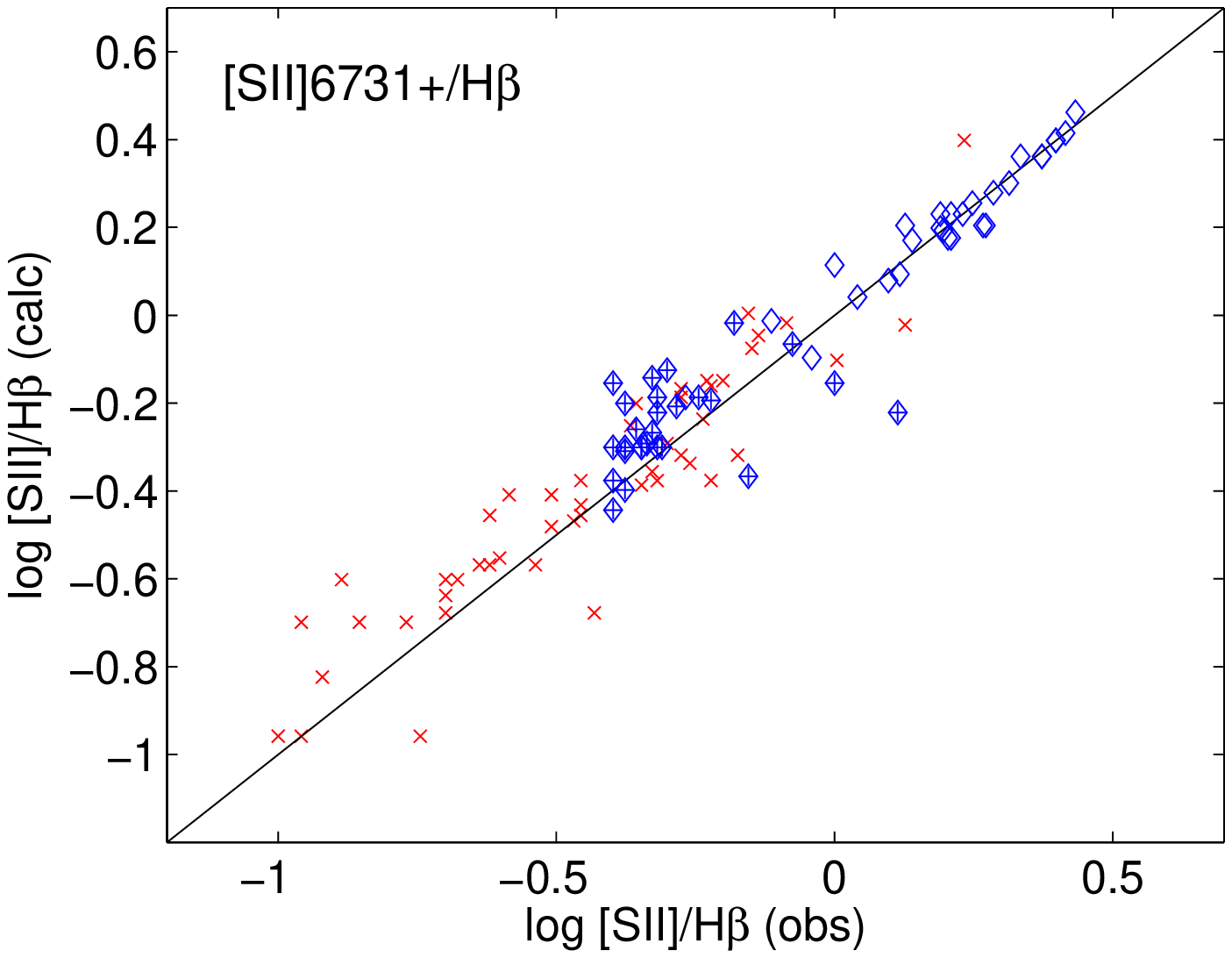}
 \caption{Observed versus calculated line ratios. Cross : Berg et al. data;
plus : Marino et al data ; diamond :  AGN galaxies (References in Table 5);
diamond+plus : Gallagher et al data.
}
\end{figure*}

\section{Modelling  the spectra}

Metallicities obtained  for galaxies  in
surveys at different redshifts were  calculated (see Contini 2016a,b, 2015, 2014 and references therein)
by the detailed modelling of the spectral lines
in the UV - optical - near IR range. Generally, the strongest lines are
[OII]3727+3729 (hereafter [OII]3727+), [OIII]5007+ 4959 (hereafter [OIII]5007+) and [NII]6584+6548 (hereafter [NII]6584+)
together with  \Hb and \Ha.
The [OIII]4363 line, which plays a dominant role in the modelling process (see Contini 2016b),   is weak and not always available.
Some spectra contain also  [NeIII]3869,  [SII]6717, [SII]6731  and   in a few cases He, Fe and Ar lines
(e.g. Hammer et al 2006).

\subsection{Line intensity calculation}

We use composite models which account for the coupled effect of photoionization and shocks
because most galaxies at high and low redshifts are the product of merging.
Photoionization and heating by a radiation source and collisional process due to shocks are
used  in the calculation of the line fluxes. The code SUMA is adopted.
The main physical properties  of the emitting gas and the element abundances are accounted for.
The input parameters such as  the shock velocity \Vs, the atomic preshock density \n0 and
the preshock magnetic field \B0 (for  all models \B0=10$^{-4}$Gauss is adopted)
define the hydrodynamical field.  They  are  used in the calculations
of the Rankine-Hugoniot equations  at the shock front and downstream and  are combined in the
compression equation which is resolved  throughout each slab of gas in order to obtain the density
profile downstream.
 The input parameters  that represent the primary radiation
from  the SB are  the effective temperature \Ts and the ionization parameter $U$.
 A pure black body radiation referring to \Ts is a poor approximation for a SB, even adopting a dominant 
spectral type (see Rigby \& Rieke 2004). However, the line ratios that are used to indicate \Ts, also depend on metallicity, 
electron temperature, density, ionization parameter, the morphology of the ionized clouds, and in particular,
they depend on the hydrodynamical field.
For an AGN, the primary radiation is the power-law radiation
flux  from the active centre $F$  in number of photons cm$^{-2}$ s$^{-1}$ eV$^{-1}$ at the Lyman limit  and
spectral indices  $\alpha_{UV}$=-1.5 and $\alpha_X$=-0.7.
  The primary radiation source
is independent  but it affects the surrounding gas.
The  region surrounding the radiation source is not considered as a
unique cloud, but as an ensemble of slabs with different thickness. 
The secondary diffuse radiation is emitted from the slabs of gas heated
by the radiation flux reaching the gas and collisionally  by the shock.
Primary  and  secondary radiation are  calculated by radiation
transfer throughout the slabs downstream.
In our model the line emitting  region throughout the galaxy covers  
 an ensemble of fragmented clouds. The geometrical
thickness of the clouds is  an input parameter of the code ($D$) which is
calculated consistently with the physical conditions and element abundances of
the emitting gas.
The fractional abundances of the ions are calculated resolving the ionization equations
for each element (H, He, C, N, O, Ne, Mg, Si, S, Ar, Cl, Fe) in each ionization level.
Then, the calculated line ratios are compared with the observed ones. The calculation process is repeated
changing  the input parameters until the observed data are reproduced by model results, within 10 percent
for the strongest line ratios and within 50 percent for the weakest ones.
  When the  data are available, the models are also constrained by the fit of the continuum spectral energy
distribution (SED).

In our previous investigations (Contini 2017, 2016a and references therein) the number of objects
selected from the different  surveys   was  relatively small
because   only the  line spectra
showing   enough significant observed line ratios  constraining the
modelling process were considered.
In the optical domain,  they are  [OII]3727+/\Hb ,
[OIII]5007+/\Hb, [NII]6584+/\Hb and \Ha/\Hb. The \Ha/\Hb line ratios are used to correct the observed data
for reddening. 
The [OII]/[OIII] corrected line ratios are used to determine the physical parameters
characteristic of the host galaxies, i.e.  the photoionizing source and, in particular, the O/H relative abundances,
because oxygen is a strong coolant
and   affects the cooling rate of the gas in the recombination regions downstream of ionization and shock fronts.
[NII]/\Hb and [SII]/\Hb are used to calculate N/H and S/H, respectively.
 Sometimes the observed \Ha/\Hb are relatively high ($>$4).  They can be found in high density gas ($>10^6$ \cm3)
where  self-absorption occurs in the Balmer lines (Osterbrock 1974). This leads to the strengthening of the \Ha ~
relatively to the other lines of the Balmer series (see also Contini 2003). Shock waves which are accounted for in the
SUMA calculations compress the gas downstream  to n/\n0$>$ 4 (the adiabatic jump) depending on \Vs.
Unfortunately \Hg ~ lines are weak and often blended with [OIII]4363 line. The 
observation error  
could reach values $>$ 50 percent. Therefore, \Hg/\Hb cannot be used to interpret the spectra properly.
 Therefore, before correcting the line ratios for reddening, 
the forbidden lines,
in paticular [OII] and [SII], are checked because the critical densities for collisional deexcitation
are $\geq$ 4$\times$10$^3$ \cm3, while  for the [OIII]5007+ lines it is  $\sim 6.5\times10^5$ \cm3.
High density gas in the host galaxy can be revealed by relatively strong 
permitted lines and abnormally low forbidden ones.
Alternatively, considering that the line fluxes are affected  by gas and dust throughout their way to Earth,
the data are reddened corrected by the \Ha/\Hb (case B) even implying errors within 20 percent.

The calculations by the code SUMA adopt an initial set of parameters  suggested by the observation evidence.
By the detailed modelling method,  the O/H and N/H relative abundances are calculated consistently
with the other elements included in the  line spectrum.
 This means that    model  parameters  are  selected
 cross-checking calculated with  observed line ratios until a fine tune is obtained for all of them.
Using  the detailed modelling of the line spectra  observed from SN and GRB hosts in our previous works, 
we have generally obtained
O/H relative abundances higher than those found
 adopting strong-line direct methods.  Higher O/H were justified by theoretical arguments considering that the 
electron  temperature and density
of the emitting gas throughout the host galaxy  are not constant, but decrease with gas recombination.
Moreover, the data are observed at Earth but calculated at the emitting nebula. The observations for nearly all
the  objects are averaged over  the whole  galaxy, so a perfect fit of the spectral line ratios is not realistic. 
Special features  are  often lost.

\subsection{Models for HII region spectra}

HII regions are common near very hot young stars. 
WD stars have temperatures close to 2 10$^5$ K.
In Table 1  the  modelling of Marino et al (2013) spectra from galaxies hosting HII regions are presented.
 Marino et al observations were obtained with the Calar Alto Legacy Integral Field Area (Califa)
by using the Potsdam Multi Aperture Spectrograph (PMAS) at CAHA 3.5 m.
In Table 1 the redshift is shown in column 2, followed  by the observed line ratios to \Hb in columns 3-6.
The observed flux in 10$^{-16}$ \erg appears in the last column.
Each row containing the data is followed by a row containing  calculation results by models mc1-mc16,
which are described in Table 2.
All the models refer to outflowing matter, except for galaxy NGC7489-007 spectrum which required an inflow model.

Berg et al (2012) MMT spectroscopic observations
of HII regions in 42 low luminosity galaxies with the Spitzer Space Telescope
that combines IRAC and Multiband Imaging Photometer (MIPS) IR imaging for 258 galaxies
in the nearest 11 Mpc of our local universe, are shown in Table 3.
The observed line ratios for each galaxy are  reported in columns  3-14. In the row   following  each dataset the
 results of modelling (models mb1-mb51)  are given. The  input parameters  are shown in Table 4, where models
mb1-mb51 are described.
Galaxy KKH37  does not appear  because the data are not sufficient to constrain the models.
For all the models adopted to reproduce   Berg et al data
the emitting clouds are inflowing towards the  starburst, as suggested by  Michalowski et al (2016).
This option was strengthened by the modelling of [OIII] 4363/\Hb line ratio, because
outflowing model results underpredicted these ratios by a factor of $\sim$3.
The calculated  star temperatures \Ts   range between 6.6$\times$10$^4$K and 1.1$\times$10$^5$K for the Marino et al
spectra and between 4.4 and 9.9$\times$ 10$^4$K for the Berg et al ones. The ionization parameter $U$ is low ($<0.1$)
for both samples.
Tables 1 and 3  and  Fig. 2 diagrams  show that calculated line ratios reproduce the observed ones within 10 percent for the
strong lines ([OII], [OIII]5007+, [NII] and roughly for [SII]), while weak lines are reproduced by $\sim$ 50 percent.
 [SIII] 6312 lines are  easily blended with  [OI]6300, 6363  and the [OIII]4363 line  with \Hg.

We compare in Fig. 3 the N/O line ratios
calculated with the strong-line direct methods by Berg et al with those calculated in the present paper.
The resulting trend suggests that  discrepancies are in average systematic.
Fig. 3 shows that the N/O ratios calculated by direct methods are shifted towards higher values,
indicating that the ratios between the (O/H)$_d$ (evaluated by direct methods)  are smaller than those
(O/H)$_m$ (calculated by the detailed modelling).  

\subsection{Calculations of  AGN spectra}

We refer to the spectra assembled and modelled by Dors et al (2017), making use of the code CLOUDY.
We have selected the galaxies   corresponding to lines uncontaminated by  broad  components,  adapted to Seyfert
2 galaxies  and corrected for reddening. 
The  observation source references  are reported in Table 5,  last column.
The observed line ratios from the AGN sample
appear in columns 3-6 of Table 5 and the best fitting  results by models ms1-ms28 are given in the next row.  
Observed line ratios and model results are compared in Fig. 2, which
  shows that the data are   satisfactorily reproduced. 
The models are described in Table 6. Note that \Vs and \n0 are higher than those calculated
for HII, SB and LGRB host galaxies, similar to those in the NLR of AGN.

We  can now compare in Fig. 3 the results obtained by a pure photoionization code  (CLOUDY) with those calculated by
 a code which refers to the 
coupled effect of shock and photoionization (SUMA). 
Fig. 3 shows that the results obtained by both the pure photoionization  code and  by the   strong-line direct method 
lead to  N/O higher 
than those obtained by models  accounting also for the shock. 
The [OIII]5007+/\Hb line ratios are reproduced by SUMA with more precision than by CLOUDY
(cf Dors et al 2017, fig. 1) because Dors et al used a constant electron density \Ne throughout the
recombination zone, whereas by SUMA  \Ne~ decreases following the cooling  rate.
Our results lead to O/H close to solar for most  AGNs and for most of the HII regions.
We will  conventionally  define   "solar"  relative abundances
 (O/H)$_{\odot }$=6.6 - 6.7 $\times$10$^{-4}$ and (N/H)$_{\odot }$= 9.$\times$ 10$^{-5}$
 (Allen 1976, Grevesse \& Sauval 1998) that were found  suitable  to local galaxy nebulae. Moreover, these values
 are included  between  those of Anders \& Grevesse (1989)
(8.5$\times$10$^{-4}$ and 1.12$\times$10$^{-4}$,respectively)   and   Asplund et al (2009)
(4.9$\times$10$^{-4}$ and 6.76$\times$10$^{-5}$, respectively ).

\subsection{Models for  TypeIa SN host galaxies}

We present in Table 7 the modelling of TypeIa SN host galaxies at z$\leq$0.03. The spectra reported by
Gallagher et al (2005) were obtained with the FAST spectrograph (Fabricant et al 1998) at the F.L. Whipple
Observatory's 1.5 m Tillinghast telescope. We have reddening corrected the line intensity fluxes.
The galaxies showing enough lines to constrain  the models were selected. For some galaxies, e.g.
IC3690, NGC2595 etc.,  the correction yields to [OII]/\Hb $\geq$ 80. Such high values  could be found adopting shock 
dominated models ($U$=0), but they would correspond to [OIII]/\Hb $>$10. The data are far lower, 
therefore we omitted them from our sample.
The observed-corrected line ratios to \Hb appear in Table 7.
Each row which refers to  the data is  followed by one which shows the  modelling results.
The models (mg1-mg27) are presented in Table 8.
Our models lead to O/H very close to solar values,
similar to those found for AGN in this z range. The N/H ratios are included between solar and 0.2 solar.
A few spectra (corresponding to NGC3627, IC4232, NGC6627 and NGC4704) were fitted by N/H =1.2-1.8 solar.

\section{Evolution of  N/O with z}

In    the present and  previous investigations the host galaxy samples were  
selected  by the  
simple criterion  that each spectrum contained enough lines to constrain the model,  without referring to specific redshifts. 
Nevertheless, Fig. 1 shows that the SN  and LGRB host domains do not entirely overlap throughout z.  
N/O ratios calculated for SN and LGRB host galaxies  peak at different z and have 
different trends throughout  the redshift. 
We suggest that  LGRB hosts are  observable at higher z because  they are
more luminous.  
In Figs. 4 and 5   the host galaxies are  distinguished in subclass objects recognizable  by
different  symbols. They  refer
to  the different types of SN, to LGRB, SGRB   and to AGN, LINER and SB   galaxies, for comparison.
We have added in the diagrams N/O, O/H and N/H model results  presented in Tables 2 and 4 for HII region 
hosts calculated
using the data of Marino et al (2013) and Berg et al (2012), respectively and for AGN  (Table  6) and Type1a SN hosts
(Table 8). 
Symbols are described in  Table 9.
In the bottom diagrams of Figs. 4 and 5  log(N/H) and log(O/H) distributions for all type of objects 
are shown as function of z.
Fig. 1 shows  that, at the state of the art, SN and LGRB N/O   domains overlap at 0.03$<$z$<$1.  
In this z range the N/O ratios for most objects are   $<$0.1  (solar N/O =0.136), except for a few objects.
N/O ratios in SN host band  peak at a lower z (Fig. 4) than  LGRB hosts (Fig. 5).
 In the upper panels  of Figs. 4 and 5   we have added the cosmic ages corresponding to the redshifts
 at which significant changes occurred in the N/O  trends.

\subsection{SN,  AGN and LINER host domains throughout the redshift}

In Fig. 4, we distinguish 4 zones  relative to the  N/O behaviour  of SN hosts and other galaxies as function of z.

1) At z $>$1.58  (zone 1) at an age $<$ 4.125 Gyr, we have significant data only for AGN.  
N/O  spans between $\sim$0.2 to $\sim$0.015,
due to low N/H and O/H  close to solar, indicating  that massive stars dominate.
(We adopt (N/H)$_{\odot}$=9.$\times$10$^{-5}$ and (O/H)$_{\odot}$=6.6$\times$10$^{-4}$).

2)  At 0.16$<$z$<$1.58 (zone 2) at an age between 4.12 and 11.7 Gyr, N/O ratios for most SN  types
  span between $>$ 0.16 and $<$0.01. Both O/H and N/H extend  towards ratios lower than solar by 
factors of $\sim$10, where the N production is primary.
Intermediate mass stars appear.
 SLSNI and other SN types are seen throughout  the   N/O band of SN hosts.

3) At   0.018$<$z$<$0.16 (zone 3) at an age between 11.7  and 13.47 Gyr, the N/O band of SN hosts 
increases rapidly  to solar values.
Both N/H and O/H reach solar values, but with different trends. 
Following Henry et al (2000), secondary N is created.

4) At z$<$0.018 (zone 4) at an age $>$13.48 Gyr the O/H and N/H ratios  in AGN are $\sim$ solar, reaching N/H
 $\sim$ 5. 10$^{-4}$  by  N secondary production at low z.
 The distribution of   N/O data in Fig. 4 for AGN and Seyfert 2 galaxies strengthens  the evidence that AGN (+LINERs) 
 extend the N/O trend of SN hosts   towards low z.
  SN Type 1a hosts fill the gap within the SN band (in particular in the lower edge) at low z, close to the AGN.

Summarizing, Fig. 4  shows that:
N/O calculated for super-luminous SN typeIb (SLSNI, Leloudas et al. 2015) host galaxies are located 
at most in the lower  edge of the SN band.
Few super-luminous SNRI  with  low O/H  are located on the  top edge of the SN host band.
 SLSN hosts appear  throughout a large range of N/O at z$\sim$0.1.
N/O  calculated for LINERs and AGN   are maximum at  z$\sim$ 0.01, then they split into two main trends
towards lower z, following the N/H trends.

\begin{figure}
\centering
\includegraphics[width=8.9cm]{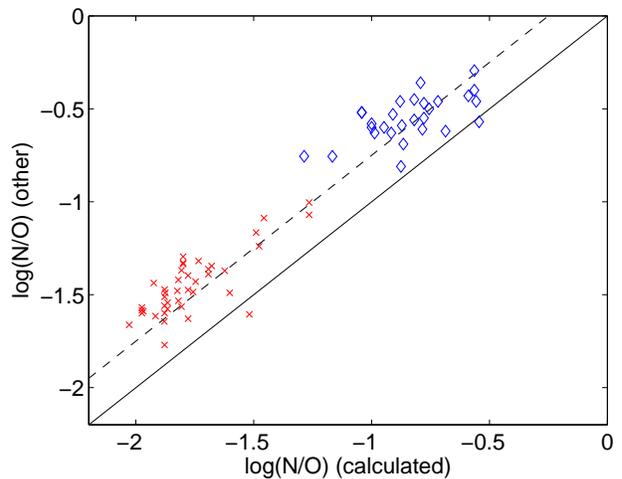}
 \caption{Comparison of N/O calculated by detailed modelling  with N/O calculated by strong-line direct methods 
by Berg et al (2012) (cross) and with the pure photoionization code CLOUDY by Dors et al (2017) (diamond)
}
\end{figure}

\subsection{LGRB, SB and HII region host domains throughout  the redshift}

In Fig. 5, we roughly distinguish 4  zones  :

1) At z between $\sim$3 and $\sim$0.4  (zone 1), at an age between of 2.17 and 9.4 Gyr, N/O calculated for SB  
are higher than for  LGRB hosts, due to higher N/H and depleted
O/H. Then,  N production is primary. Intermediate  massive stars dominate in the SB host sample. 
On the other hand, O/H ratios are nearly solar for LGRB hosts and N/H  lower than solar (secondary N production
is starting), 
confirming  that massive stars dominate in the LGRB hosts;
 The primary/secondary behaviour of nitrogen was well explained by Vila-Costas \& Edmunds (1993, fig. 4).
SB host galaxies show a large dispersion of O/H ratios between solar and 0.24 solar.

2) At z between 0.4 and 0.158 (zone 2), at an age between 9.4  and 11.73 Gyr, 
a rapid fall of N/O calculated for both SB and LGRB hosts  is seen also  for N/H and O/H (bottom diagram).
The decrease of O/H for  the SB started already in zone 1. For LGRB hosts it is sharp on the upper bound distribution 
due to the results obtained for LGRB modelling the Niino et al (2016) survey host galaxies.
The minimum at 0.1$<$z$<$0.4 which corresponds to a similar minimum of N/H was explained by Contini (2017)
as due to the  merging process which implies inclusion of low  N/H and O/H matter. 
The minimum at log(z)$\sim$-0.8  corresponds to N/O$\sim$0.025 (0.18 solar).
Comparing with Contini (2017, fig. 2) at this redshift N/H shows a minimum of 10$^{-5}$ while
O/H = 5 10$^{-4}$ is recovering from a minimum of 4.6 10$^{-4}$ at log(z) $\geq$-1, 
perhaps indicating that massive stars are formed following the merging process.

3) At z between 0.158 to 0.063 (zone 3), at an age between 11.73 and 12.87 Gyr,  the LGRB hosts 
regain solar O/H and N/H.
 Massive and intermediate mass stars coexist.

4) At z$<$0.063 (zone 4) :  N/H  data for SB hosts seem  to increase towards lower z,   
reaching  also higher than solar values.
On the other hand,  the few  LGRB hosts show O/H and N/H  ratios merging with those calculated for the HII regions.
The latter show N/H values scattered around $\sim$ 0.1 solar and O/H $\leq$solar (Tables 2 and 4), 
leading to a slowly decreasing
and large spread of N/O ratios  towards low z.

Summarizing, in zone 1  intermediate mass stars dominate for SB but not for LGRB, where massive stars prevail.
In zone 2 both N/H and O/H decrease indicating merging.
In zone 3 the poor  production of secondary nitrogen leads N/H in LGRB   towards $\sim$ 0.1  solar and O/H solar,
and   low N/O.
The lowest N/H ratios  characterize SGRB host galaxies.

\section{N/O versus metallicity}

\begin{figure*}
\centering
\includegraphics[width=12.6cm]{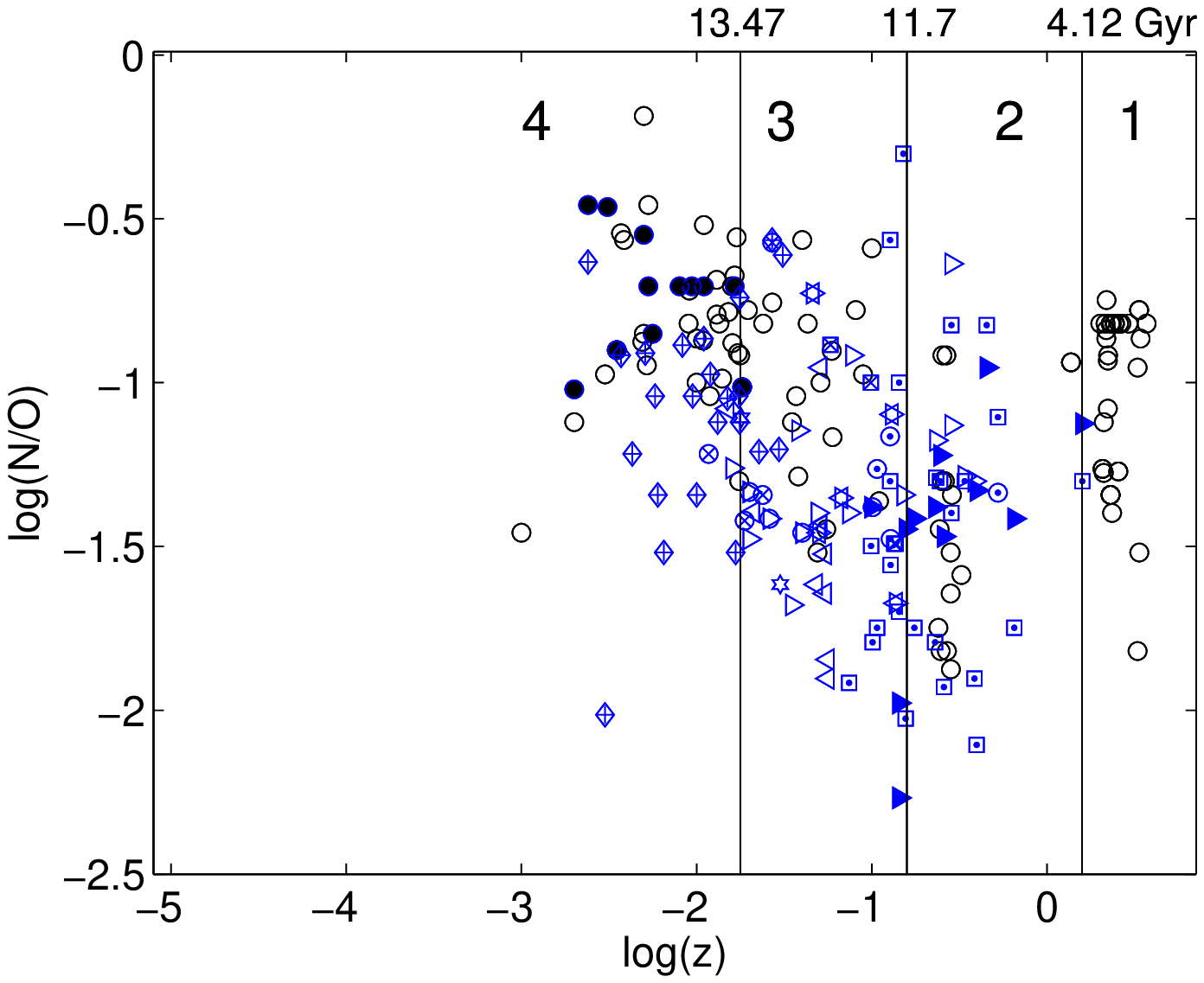}
\includegraphics[width=12.6cm]{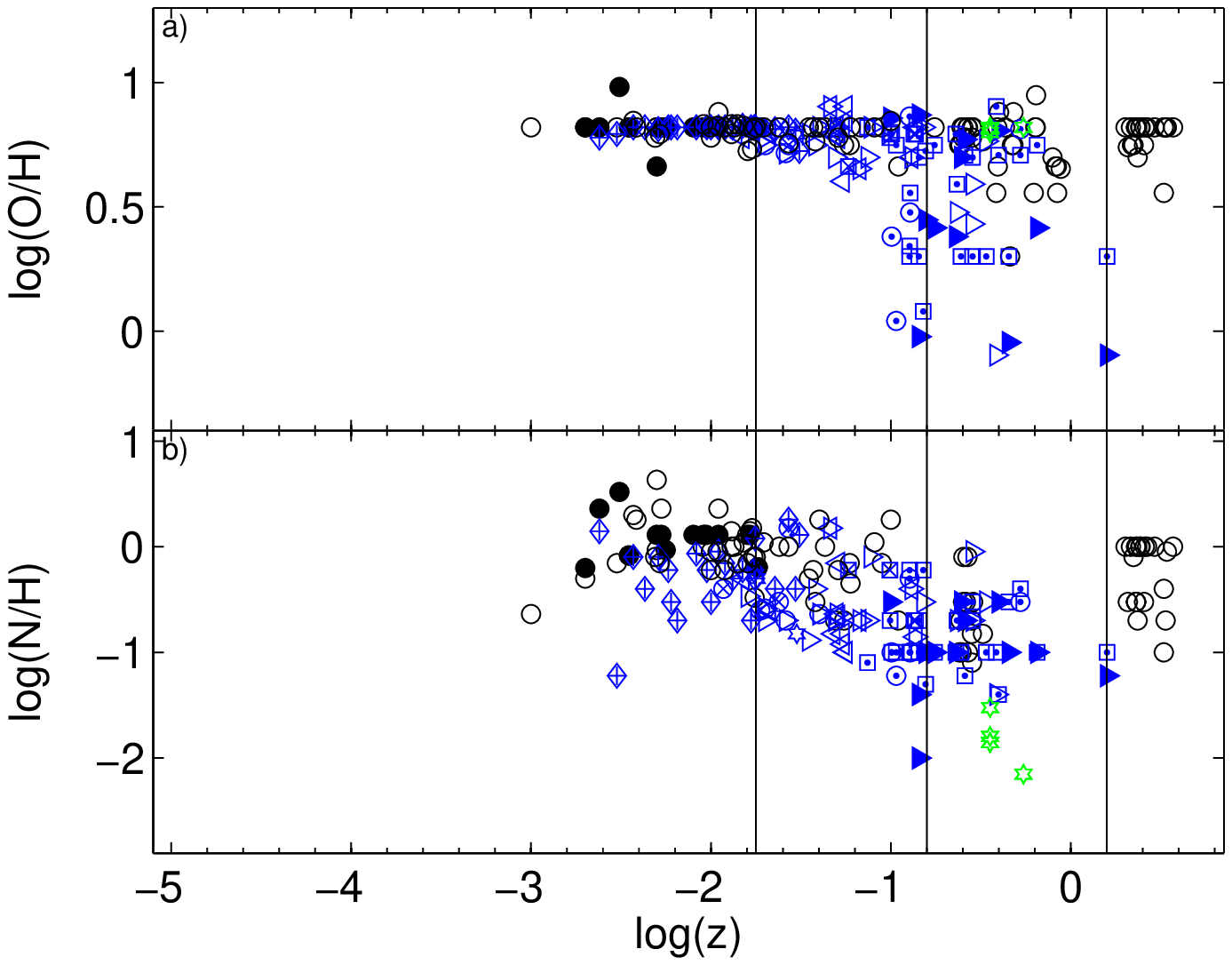}
 \caption{
Top : distribution of N/O as function of the redshift
for SN hosts, AGN and LINERs;
bottom : distribution of  O/H in 10$^{-4}$ units  and N/H in 10$^{-4}$ units
as function of the redshift for SN hosts, AGN and LINER galaxies.
Symbols are  explained in Table 9.
}
\end{figure*}

\begin{figure*}
\includegraphics[width=12.6cm]{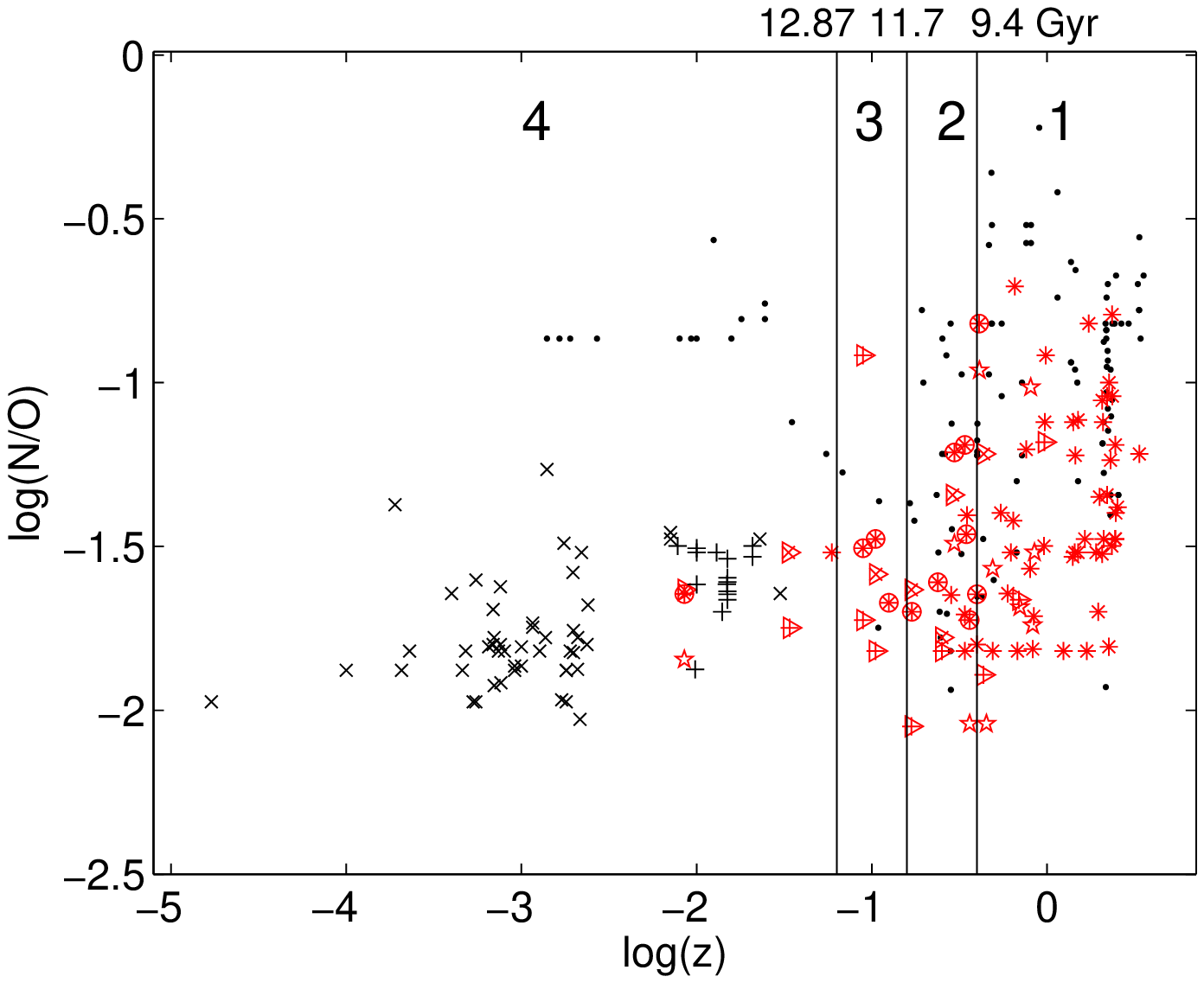}
\includegraphics[width=12.6cm]{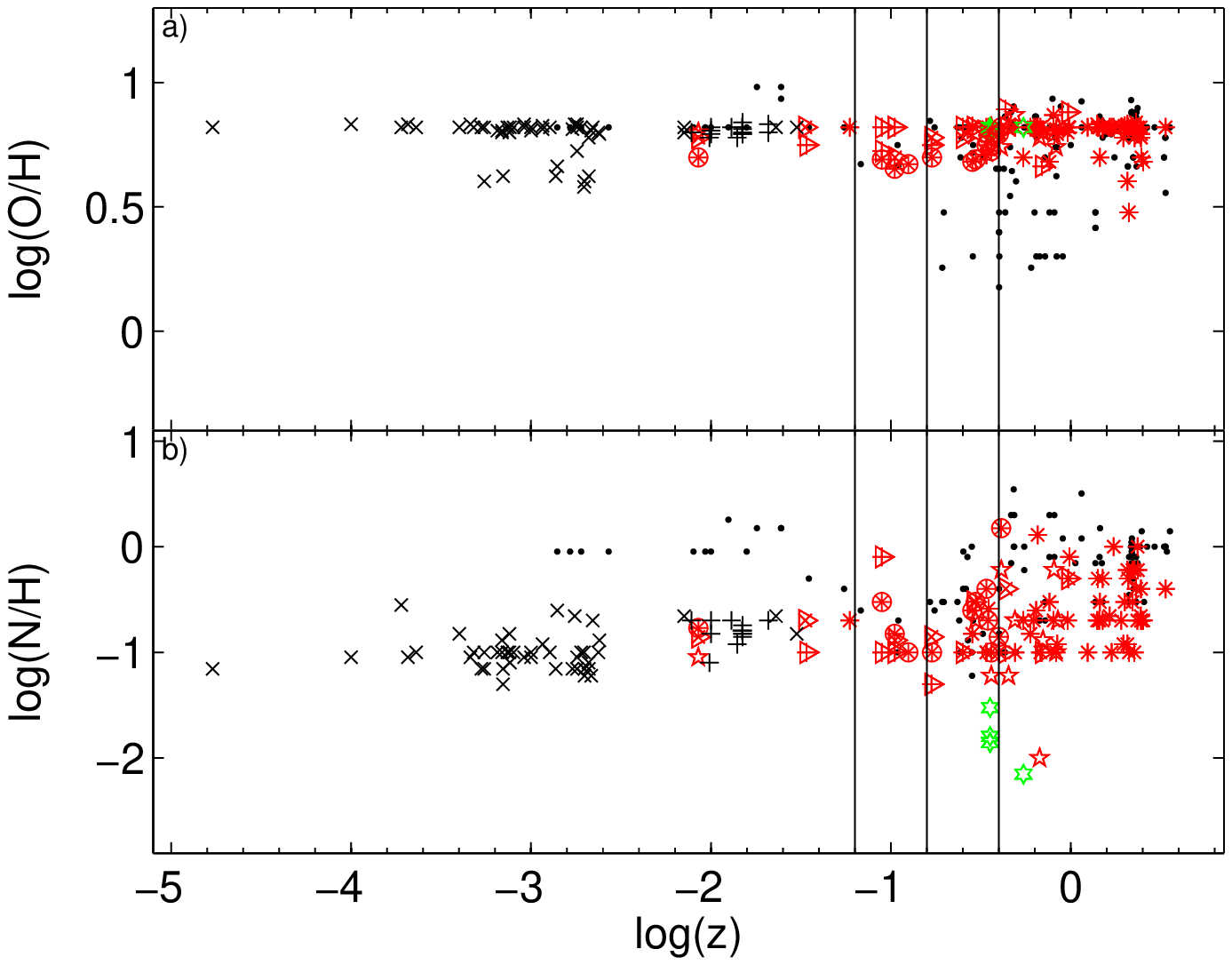}
 \caption{
The same as  in Fig. 4 for GRB,  SB and HII region  host galaxies.
}
\end{figure*}

\begin{table}
\centering
\caption{Symbols in Figs. 4 and 5}
\begin{tabular}{llcl} \hline  \hline
\ symbol   &object                       & Ref. \\ \hline
\  encircled dot&  SLSNR hosts           & (1) \\
\  triangles at z$\geq$0.1 & SLSNII hosts & (2)\\
\  filled triangles & SLSNI hosts         & (3)\\
\  square+dot & SD  models for SLSN hosts & (4)\\
\  square+cross& Type Ic host central     & (5)\\
\  circle+cross & Type Ic host at SN positions & (6) \\
\  triangles  at z$\leq$ 0.1 & SN Ib host & (7)    \\
\  encircled triangles  & SN IIb hosts    & (8) \\
\  opposite triangles    & SN Ic hosts    & (9) \\
\  double triangles      & SN IcBL hosts  & (10) \\
\  hexagrams               & SN Ibc        & (11)  \\
\  asterisks &GRB hosts                   & (12) \\
\  triangle+cross    & LGRB hosts         & (13)  \\
\  pentagrams  & LGRB different hosts     &  (14)  \\
\  triangle +plus& LGRB hosts with WR stars& (15) \\
\  encircled asterisks &LGRB at low z     & (16)\\
\   hexagrams (green) & SGRB hosts                & (17)\\
\  diamonds & other  galaxies              & (18) \\
\  dots & starburst galaxies              & (19,25)\\
\  circles & AGN                          & (20,21,25)  \\
\  filled circles & LINER                 & (22) \\
\  plus &  low-luminosity nearby galaxies & (23)  \\
\  cross& HII regions in local galaxies   & (24)  \\ 
\  diamonds+plus& TypeI SN hosts          & (26)  \\ \hline
\end{tabular}

 (1), (2), (3), (4) (Leloudas et al);
(5), (6) (Modjaz et al);
(7), (8), (9), (10), (11) (Sanders et al);
(12) (Kr\"{u}hler et al);
(13) (Savaglio et al);
(14) (Contini 2016a, table 8);
(15) (Han et al);
(16) (Niino et al);
(17) (de Ugarte Postigo et al);
(18), (19), (20) (Contini 2014 and references therein); 
(21) (Koski 1978, Cohen 1983, Kraemer et al 1994, Dopita et al 2015;
(22) (Contini 1997);
(23) (Marino et al 2013);
(24) (Berg   et al 2012);
(25) (Contini 2016b);
(26) (Gallagher et al 2005).

\end{table}

We have noticed in the previous sections that, relatively to the N/O distribution, 
  HII regions can be considered as the  continuation of the LGRB host distribution
at low z.  
 Comparing the physical parameters presented  in Tables 2 and 4 for HII regions with those presented 
for LGRB, SN and SB (Contini 2017 and references therein)  slightly lower  \Vs, \n0 and $U$ were 
found for HII regions,  but similar \Ts, 
indicating that the  photoionizing source of the host galaxy gas is roughly the same.
 SB galaxies are distributed along the whole z range. However, most of them
 at high z show  N/O ratios  higher than  in LGRB hosts,
due to higher  N/H and lower O/H ratios. 
It is generally considered that the SN and GRB hosts are SB dominated.
However,  SB and LGRB host characteristics  diverge at high z.
Moreover, 
 Figs. 4 and 5 show that the N/O ratios calculated for SN and LGRB host galaxies
follow roughly the trends of N/H as function of z. This indicates that N and O ejections
originating from the bursts have different  histories in times.

To have a wider picture of these features  we  now investigate whether  the  models calculated 
for the chemical evolution of starburst galaxies
by  MC02 could explain the behaviour  of N/O as function of O/H for
 the different galaxy types (SN, GRB hosts and other galaxies).
MC02 claim that the variation of N/O at constant metallicity may originate from a time delay between the 
release of oxygen
and that of nitrogen in the ISM. The delayed-release model assumes that star formation is an intermittent process 
in galaxies and the delayed release of nitrogen produced in long lived low mass stars compared to oxygen 
produced in massive short-lived stars should be considered.
Multi-burst scenarios were found for massive starburst nucleus galaxies.

In Figs. 6 some significant model results from the calculations of  MC02 (their fig. 1) about
continuous star formation and bursting star formation,  are schematically overlied   
 upon the results obtained by the detailed modelling of the spectra in previous sections.
Log(N/O) is shown as function of log(O/H)+12.
We have  reported in Fig. 6 MC02 (fig. 1, left diagram)   lower and higher bounds
of continuous star formation models, including the low N/O-O/H left-bottom region, by thick and thin
solid lines, respectively. The models merge in the top-right region at high log(N/O) and high log(O/H).
The  thick-solid  line in Figs. 6 refers to models with an infalling time $\tau$$_f$= 4 Gyr and a star formation
 timescale $\tau$$_{SF}$=2 Gyr, while the   thin line refers to $\tau_f$=7 Gyr and $\tau_{SF}$=20 Gyr.
The  dot-dashed lines  circumscribe the regions of   star-bursting  formation models
(MC02, fig. 1, right diagram) at low and high log(O/H). 
We report our results on the basis of Figs. 4 and 5, i.e.  distinguishing between  LGRB, SB and HII regions
in Fig. 6 (bottom diagram) and SN and AGN hosts (top diagram).
 Symbols are given in Table 9.
  In the LGRB sample,   the  dichotomy  noticed by MC02 between  galaxies with 
log(O/H)+12 higher or lower than 8.5,
is   roughly evident. Some LGRB host galaxies appear  near continuous star formation models,  most of them are
located  within the 
star-bursting formation domain and beyond, where also some SN hosts can be seen.
The low O/H - N/O region  refers mostly to dwarf irregular or metal poor UV-selected galaxies.  
Calculated N/O for HII region do not appear in this region.
At 12+log(O/H) $>$ 8.6 many different objects are nested close to O/H solar  with  N/O
ranging between the maximum shown by the  starburst galaxies  and AGN and  the minimum shown  by the  HII regions. 
SGRB are located at  extremely low N/O, due to low N/H.
HII region and SB host galaxies splits at the same metallicity in  low and high N/O, respectively.

Relatively high N/O calculated in some  SN hosts, SB and AGN  are  explained by star-bursting  models.
In general,  N/O in SN hosts (Fig. 6 top)  do not follow MC02 model results at low O/H.
Therefore, the SN host  characteristics are  different from those of the  HII region and LGRB hosts.

\begin{figure}
\includegraphics[width=8.6cm]{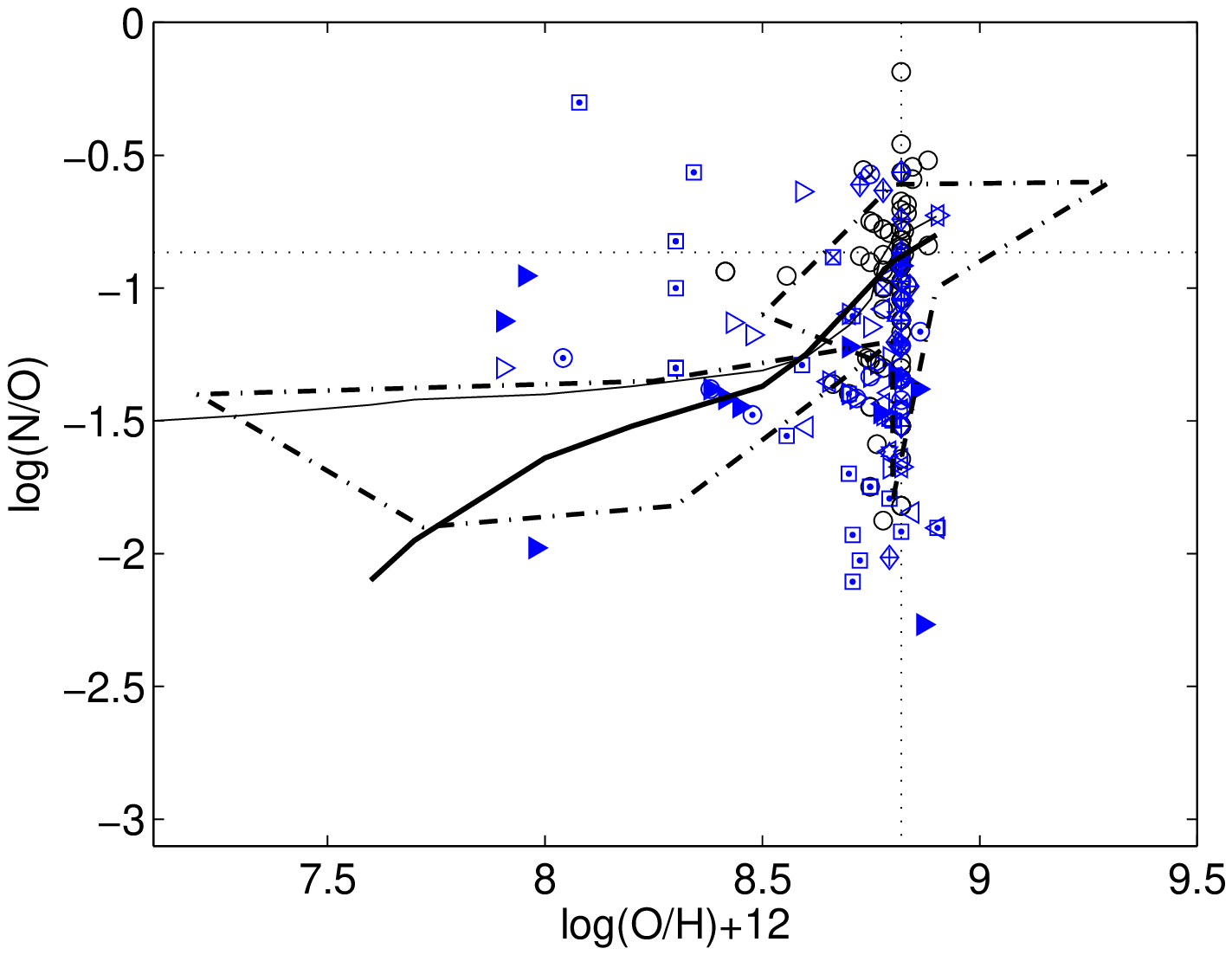}
\includegraphics[width=8.6cm]{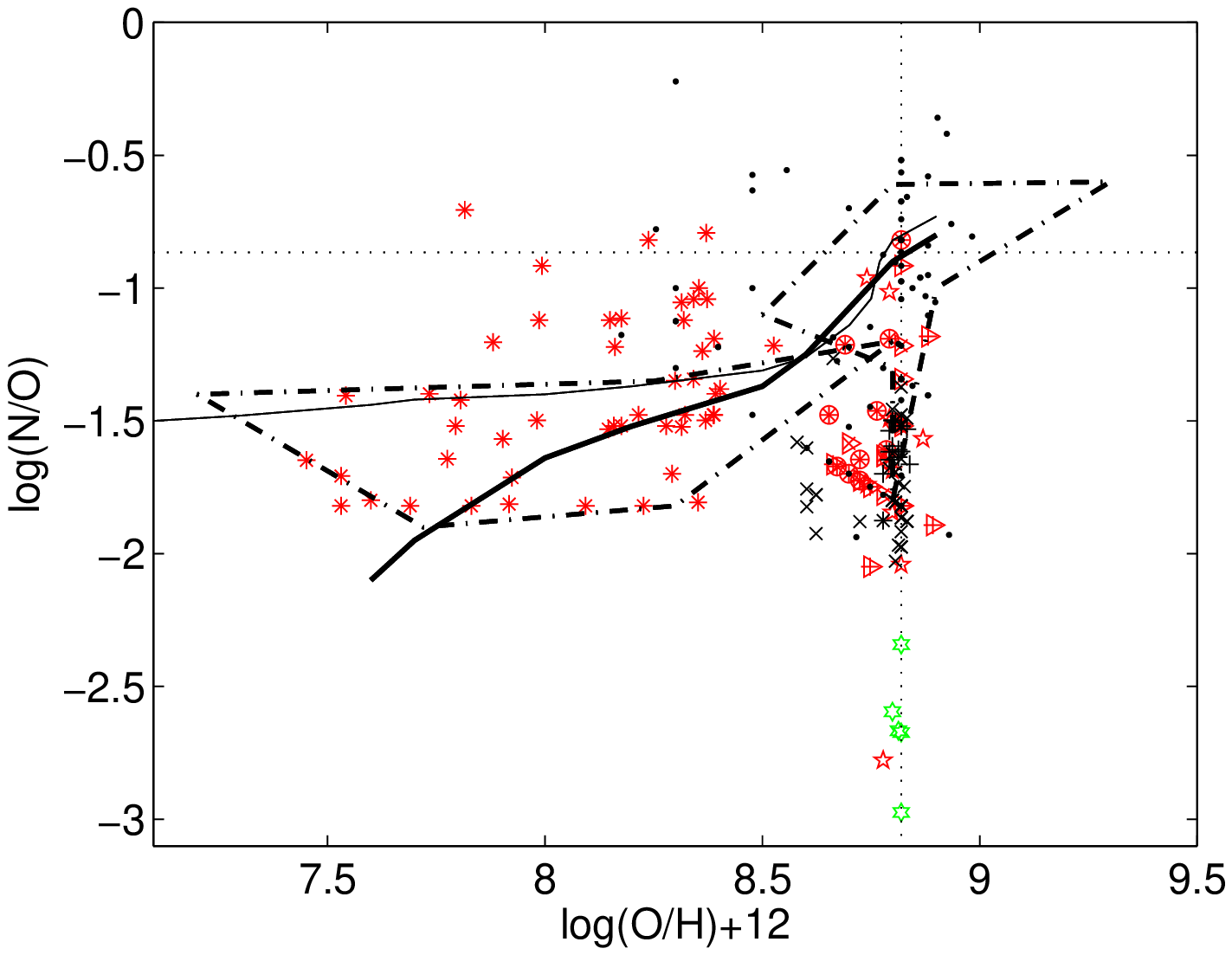}
 \caption{Top : log (N/O) versus log(O/H)+12 for different types of SN and AGN host galaxies.
 Bottom  :  the same as top diagram for different types of GRB,  SB and HII region host galaxies.
 Symbols are given in Table 9.
}
\end{figure}

\begin{figure}
\includegraphics[width=8.6cm]{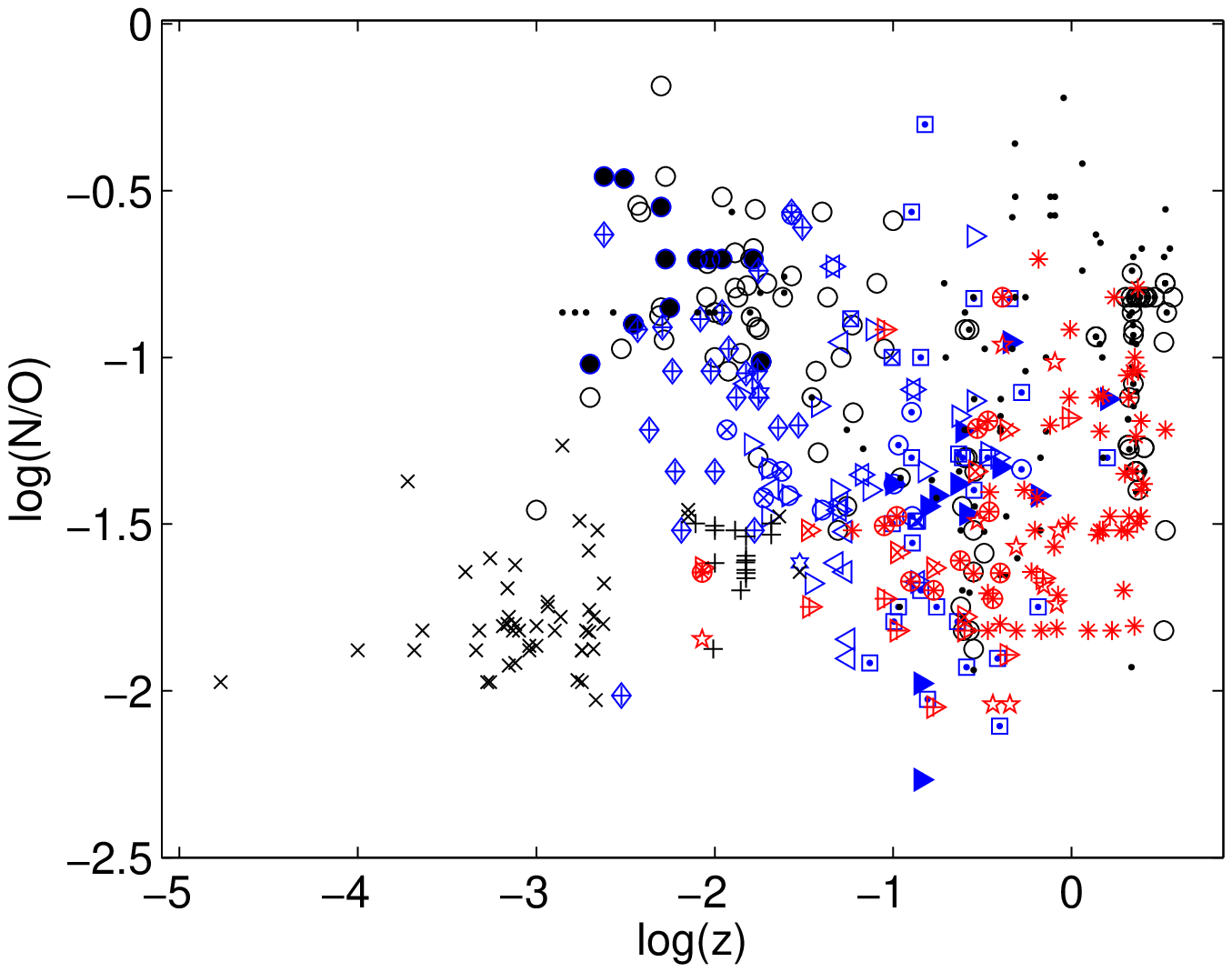}
\includegraphics[width=8.6cm]{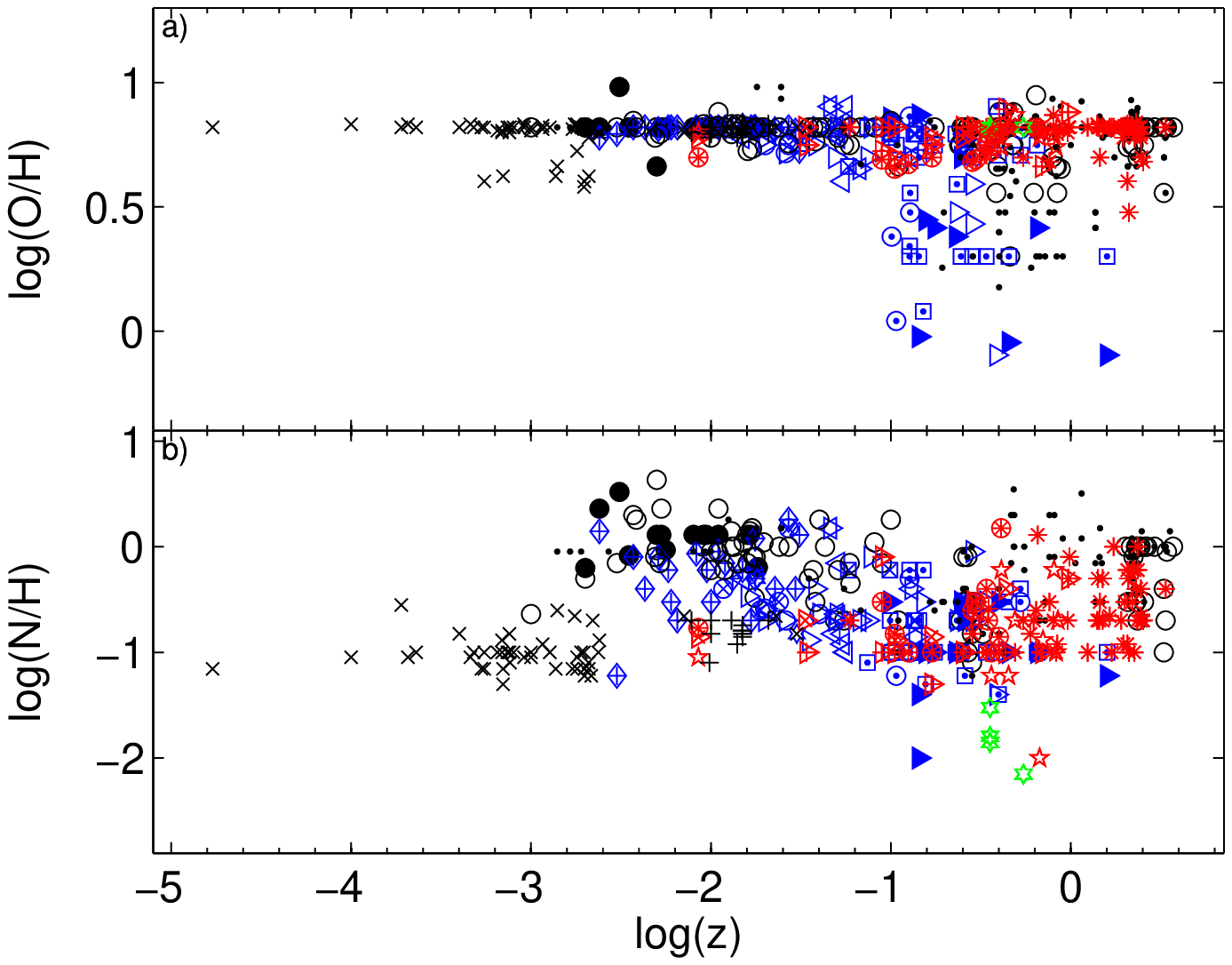}
\caption{Top: distribution of N/O as function of the redshift for SN host galaxies (squares+filled diamonds),
long GRB hosts (asterisks), short GRB hosts (stars),
  SB (dots), AGN (open circles) and LINERs  (filled circles) both local and at higher z.
Bottom : distribution of O/H and N/H as function of the redshift
 for SN, long GRB, short GRB, AGN, LINERs and SB hosts. Symbols as in top
diagram;
plus and cross : HII region galaxies nearby and local, respectively.
}
\end{figure}

\section{Concluding remarks}

The distribution of the N/O abundance ratios    over the full redshift range considered here   is calculated by the detailed modelling 
of  line spectra in order to   compare the   characteristic metallicities
in various  galaxy types, in particular SN and LGRB hosts at z$\leq$3.5 (age $\geq$1.823 Gyr).
 SN and LGRB host galaxy  spectra  reported in this paper were  selected  among those 
showing enough significant lines to constrain the models.  
The available  data  for galaxies at z$>$3.5 are  too few to draw reliable conclusions. 
Our results are summarized in Fig. 7.
We have found that the N/O abundance ratios calculated to fit the line ratios  peak at different z, 
higher for LGRB than for SN hosts,
depending perhaps on  larger luminosities  in  LGRB hosts.
They also  show different trends as function of z.

The N/O ratios  calculated  in SN hosts  appear  within a band  which increases from z $\sim$ 0.63 (6 Gyr) 
towards z$\sim$0.01 (13.58 Gyr),
following secondary N formation.  O/H rapidly attains solar values at z $\sim$0.1 (12.41 Gyr) in almost all objects, 
while at z$>$0.1 the N/O ratios   spread  within two orders of magnitude depending  on   the  O/H distribution. 
AGN and LINERs  
follow the SN N/O trend and  extend it towards  lower z, reaching N/O higher than solar.

 SN hosts and AGN  at log(O/H)+12$\leq$8.5 
are  not well   explained by the MC02 models.
O/H  in most SN hosts are nested  close to solar   within the N/O   star-bursting formation  area,  
but N/O in SN hosts   are lower
than in AGN,   due to   different time release of N/H and O/H.

The upper bound of the N/O trend  calculated for LGRB host galaxies  decreases between z=3 
and z=0.1. They  spread at z$>$0.3 (10.27 Gyr)  due to the N/H distribution, which  ranges between
solar and 0.1 solar.
Massive and intermediate mass galaxies most probably coexist in this z range.
At z$<$ 0.1, N/O in the few LGRB hosts  included in the sample follow the slowly decreasing trend  calculated 
in nearby HII regions, while
N/O calculated   in SB galaxies, which coexist with LGRB hosts  at z$>$0.1,   are in average
higher than solar, due to higher N/H and lower O/H. N corresponds to primary production.

 Comparing LGRB hosts with SB galaxies  in the N/O versus metallicity diagram (Fig. 6), N/O in SB galaxies 
at log(O/H)+12 between  8.5 and 8.82  are distributed on a large N/O range.
 N/O in LGRB hosts with solar O/H and N/O $<$0.1 solar  are   close to those calculated 
in HII regions  which reach  N/O$\sim$ 0.01.
 The lowest N/O  are  calculated in SGRB hosts, which  are, however, represented by a few objects.
LGRB hosts  at 12+log(O/H)$<$8.5 are included among  star-bursting formation models. 
Some  objects follow the  continuous star formation trend
referring to models with an infalling time $\tau$$_f$= 4 Gyr and a star formation
 timescale $\tau$$_{SF}$=2 Gyr.

More data, in particular at high z, will confirm and complete our results.

\section*{Acknowledgements}
I am very grateful to the referee for many critical suggestions which substantially
improved the presentation of the paper.

\section*{References}

\def\ref{\par\noindent\hangindent 18pt}

\ref Allen, C.W. 1976 Astrophysical Quantities, London: Athlone (3rd edition)
\ref Anders, E., Grevesse, N. 1989, Geochimica et Cosmochimica Acta, 53, 197
\ref Asplund, M., Grevesse, N., Sauval, A.J., Scott, P. 2009, ARAA, 47, 481
\ref Berg, D.A. et al 2012, ApJ, 754, 98
\ref Cohen, R.D. 1983, ApJ, 273, 489
\ref Contini, M. 2017, MNRAS, 466, 2787
\ref Contini, M. 2016a, MNRAS, 460, 3232 
\ref Contini, M. 2016b, MNRAS, 461, 2374 
\ref Contini, M. 2015, MNRAS, 452, 3795
\ref Contini, M. 2014, A\&A, 564, 19
\ref Contini, M. 1997, A\&A, 323, 71
\ref Contini, T., Treyer, M.-A., Sullivan, M., Ellis, R.S. 2002, MNRAS, 330, 75
\ref de Ugarte Postigo, A. et al 2014, A\&A, 563, 62
\ref Dopita, M.M. et al. 2015, ApJS, 217, 12
\ref Dors, O., Arellano-C\`{o}rdova, K., Cardaci, M., Hagele, G. 2017, MNRAS in press
\ref Edmunds, M.G.  \& Pagel, B.E.J. 1978, MNRAS, 185, 777
\ref Edmunds, M.G. 1990, MNRAS, 246, 678
\ref Fabricant, D., Cheimets, P., Caldwell, N., Geery, J. 1998, PASP, 110, 79
\ref Gallagher, J.S., Garnavich, P.M., Berling, P., Challis, P., Jha, S., Kirshner, R.P. 2005, ApJ,634, 210
\ref Grevesse, N., Sauval, A.J. 1998 SSRv, 85, 161
\ref Hammer, F. et al 2006, A\&A, 454, 103
\ref Han, X. H., Hammer, F., Liang, Y. C., Flores, H., Rodrigues, M., Hou, J. L., Wei, J. Y.
 2010, A\&A, 514, 24	
\ref Henry, R.B.C., Edmund, M.G. \& K\"{o}ppen, J. 2000, ApJ, 541, 660
\ref Kewley, L.J.\& Dopita, M.M. 2002, ApJS, 142, 35
\ref K\"{o}ppen, J. \& Hensler, G. 2005, A\&A, 434, 531
\ref Koski, A. 1978, ApJ, 223, 56
\ref Kraemer, S.B., Wu, C.-C., Crenshaw, D.M.,Harrington, J.P. 1994, ApJ, 435, 171
\ref Kr\"{u}hler, T. et al 2015 A\&A, 581, 125
\ref Leloudas, G. et al 2015, MNRAS, 574, A61
\ref Marino, R.A. et al 2013, A\&A, 559, 114
\ref Michalowski, M.J. et al 2016 arXiv:1609.01742
\ref Modjaz, M. et al 2008, AJ, 135, 1136
\ref Mouhcine, M., Contini, T. 2002, A\&A, 389, 106
\ref Niino, Y. et al 2016 Publ. Astron. Soc. Japan,  ArXiv:1606.01983
\ref Osterbrock, D. E. 1974 in
 Astrophysics of gaseous nebulae, San Francisco, W. H. Freeman and Co., 1974. 
\ref P\'{e}rez-Montero, E. et al. 2013, A\&A, 549, A25
\ref P\'{e}rez-Montero, E. \& Contini, T. 2009, MNRAS, 398, 949
\ref Pilyugin, L.S. 1993, A\&A, 277, 42
\ref Piranomonte, S. et al 2015, MNRAS, 452, 3293
\ref Sanders, N.E. et al. 2012 ApJ, 758, 132
\ref Savaglio, S., Glazerbrook, K., Le Borgne, D. 2009, ApJ, 691, 182
\ref Sollerman J.; \"{O}stlin, G.; Fynbo, J. P. U.; Hjorth, J.; Fruchter, A.; Pedersen, K.
        2005, NewA, 11, 103
\ref Tremonti, C.A. et al. 2004, ApJ, 613, 898 
\ref Vergani, S.D. et al 2011 A\&A 535, A127
\ref Vila-Costas, M.B. \& Edmunds, M.G. 1993, MNRAS, 265, 199
\ref Woosley, S.E. 1993, ApJ, 405, 273

\end{document}